\pdfoutput=1
\documentclass[fleqn,usenatbib]{mnras}

\usepackage{mathptmx}
\usepackage{pdflscape}

\usepackage[T1]{fontenc}
\usepackage{ae,aecompl,pdflscape}

\usepackage{ulem}
\usepackage{graphicx}	
\usepackage{amsmath}	
\usepackage{amssymb}	
\usepackage{subfig}
\usepackage{float}
\usepackage{epstopdf}
\usepackage{multirow}

\title[The first 62 AGN in MaNGA - I: characterization and control sample]{The first 62 AGN observed with SDSS-IV MaNGA - I: their characterization and definition of a control sample}

\author[S. B. Rembold et al.]{Sandro B. Rembold,$^{1,3}$\thanks{E-mail: sandro.rembold@ufsm.br}
J\'aderson Shimoia,$^{2,3}$
Thaisa Storchi-Bergmann,$^{2,3}$
\newauthor Rog\'erio Riffel,$^{2,3}$
Rogemar A. Riffel,$^{1,3}$
N\'icolas D. Mallmann,$^{2,3}$
\newauthor Jana\'ina C. do Nascimento$^{2,3}$,
Thales N. Moreira,$^{1,3}$
Gabriele S. Ilha,$^{1,3}$
\newauthor Alice D. Machado,$^{1,3}$
Rafael Cirolini,$^{1,3}$
Luiz N. da Costa,$^{4,3}$, Marcio A. G. Maia,$^{4,3}$
\newauthor Bas\'ilio X. Santiago,$^{2,3}$
Donald P. Schneider,$^{5,6}$ Dominika Wylezalek,$^{7}$
\newauthor Dmitry Bizyaev,$^{8,9}$
Kaike Pan$^{8}$, and
Francisco M\"uller-S\'anchez$^{10}$
\\
$^{1}$Departamento de F\'isica, CCNE, Universidade Federal de Santa Maria, 97105-900, Santa Maria, RS, Brazil\\
$^{2}$Departamento de F\'isica, IF, Universidade Federal do Rio Grande do Sul, CP 15051, 91501-970, Porto Alegre, RS, Brazil\\
$^{3}$Laborat\'orio Interinstitucional de e-Astronomia - LIneA, Rua Gal. Jos\'e Cristino 77, Rio de Janeiro, RJ - 20921-400, Brazil\\
$^{4}$Observat\'orio Nacional - MCT, Rua General Jos\'e Cristino 77, Rio de Janeiro, RJ - 20921-400, Brazil\\
$^{5}$Institute for Gravitation and the Cosmos, The Pennsylvania State University, University Park, PA 16802, USA\\
$^{6}$Department of Astronomy and Astrophysics, The Pennsylvania State University, University Park, PA 16802, USA\\
$^{7}$Center for Astrophysical Sciences, Department of Physics and Astronomy, Johns Hopkins University, 3400 North Charles Street,\\
Baltimore, MD 21218, USA\\
$^{8}$Apache Point Observatory, P.O. Box 59, Sunspot, NM 88349, USA\\
$^{9}$Sternberg Astronomical Institute, Moscow State University, Moscow, Russia\\
$^{10}$Center for Astrophysics and Space Astronomy, Department of Astrophysical and Planetary Sciences, University of Colorado,\\
389 UCB, Boulder, CO 80309-0389, USA
}

\date{Accepted XXX. Received YYY; in original form ZZZ}

\pubyear{2017}

\begin{document}
\label{firstpage}
\pagerange{\pageref{firstpage}--\pageref{lastpage}}
\maketitle

\begin{abstract}
We report the characterization of the first 62 MaNGA Active Galactic Nuclei (AGN) hosts in the Fifth Product Launch (MPL-5)  and the definition of a control sample of non-active galaxies. This control sample -- comprising two galaxies for each AGN -- was selected in order to match the AGN hosts in terms of stellar mass, redshift, visual morphology and inclination. The stellar masses are in the range $9.4<\mbox{log}\left(M/M_\odot\right)<11.5$, and most objects have redshifts $\leq 0.08$. The AGN sample is mostly comprised of low-luminosity AGN, with only 17 nuclei with $L(\rm{[OIII]}\lambda 5007\AA) \ge 3.8\times 10^{40}\,\mbox{erg}\,\mbox{s}^{-1}$ (that we call ``strong AGN''). 
The stellar population of the control sample galaxies within the inner 1--3 kpc is dominated by the old ($\sim$ 4 -- 13\,Gyr) age component, with a small contribution of intermediate age
($\sim$640--940 Myr) and young stars ($\leq 40$\,Myr) to the total light at 5700\AA. While the weaker AGN show a similar age distribution to that of the control galaxies, the strong AGN show an increased contribution of younger stars and a decreased contribution of older stars. Examining the relationship between the AGN stellar population properties and $L(\rm{[OIII]})$, we find that with increasing $L(\rm{[OIII]})$, the AGN exhibit a decreasing contribution from the oldest ($>$4\,Gyr) stellar population relative to control galaxies, but have an increasing contribution from the younger components with ages $\sim$40\,Myr. We also find a correlation of the mean age differences (AGN - control) with $L(\rm{[OIII]})$, in the sense that more luminous AGN are younger than the control objects, while the low-luminosity AGN seem to be older. These results support a connection between the growth of the galaxy bulge via formation of new stars and the growth of the Supermassive Black Hole via matter accretion in the AGN phase.

\end{abstract}




\section{Introduction}

The MaNGA (Mapping Nearby Galaxies at the Apache Point Observatory) survey,  a core program of the fourth-generation Sloan Digital Sky Survey (SDSS-IV), operating between 2014 and 2020, will deliver optical integral field spectroscopic observations of $\sim$ 10,000 galaxies. It has been conceived in order to produce homogeneous spectroscopic information of these galaxies out to a radial position of at least $1.5\,r_e$ (effective radius). An overview of the main science objectives and survey design is presented in \citet{bundy15}. The integral field unit design and performance are described in \citet{drory15}. Details about the observing strategy in order to ensure spectrophotometric accuracy are given in \citet{law15}. The spectrophotometric calibration technique is described in \citet{yan16}.

Among the 10,000 galaxies, there are expected to be $\sim$\,300 galaxies hosting Active Galactic Nuclei (AGN). One of the primary goals of the MaNGA survey is to explore the relation between the AGN and their host galaxies. This investigation
will be accomplished via the mapping of the ionized and neutral gas kinematics, searching in particular for outflows and investigating the corresponding feedback effects on the host galaxy \citep{zakamskagreene2014}.
The empirical relationship between the mass of the central Supermassive Black Hole (SMBH) in AGNs and the stellar velocity dispersion of the spheroidal component of galaxies (the $M_\bullet - \sigma$ relationship) suggests that the growth of a galaxy (due to star formation) and its central SMBH (due to gas accretion) may be coupled \citep[e.g.][]{ferrarese00,gebhardt00}, although some non-causal explanations have been also recently proposed \citep[e.g.][]{peng07, jahnke11}. The observed $M_\bullet-\sigma$ relationship has been explained as being originated by the AGN feeding and feedback processes that couple the growth of the SMBHs and their host galaxies \citep{ferrarese05,somerville08,kormendy13}. Indeed, cosmological simulations suggest that AGN plays a fundamental role in the evolution of its host galaxy \citep{dimateo05,springel05,bower06}, as while the SMBH evolves together with the galaxy, it is fed by surrounding material and periodically produces gas ejections that retard the growth of the galaxy by preventing the accretion of extragalactic gas in these active phases \citep[e.g.][]{nemmen07,fabian12,terrazas16}. 

AGN feeding and feedback effects are related to the stellar population of the host galaxy. Previous studies have suggested that the feeding, leading to the growth of the central SMBH, appears to be related to recent
episodes of star formation in the circumnuclear region \citep[e.g.][]{heckman97,sb01,davies07,hickox14,Diamond-Stainic,Esquej}.  These studies found an excess of young to intermediate age stars in the inner few hundred parsecs of AGN hosts when compared to non-active galaxies and support the existence of an AGN-Starburst connection \citep{perry85,terlevich85,norman88}. This connection can be understood, as both star formation and nuclear activity may be fed by gas inflows towards de center or, alternatively the central AGN may be triggered due to mass loss from evolving stars \citep[e.g.][]{wild10}.  The AGN feedback, both via radiative and kinetic power, may, in turn, also affect the stellar populations of the host galaxy in the vicinity of the AGN by quenching star formation \citep[e.g.][]{fabian12,dubois13,ishibashi14,pontzen17,xie17}.

The present study is the first of a series of papers in which we will use the MANGA data cubes to obtain the properties of AGN and a  matched sample of inactive galaxies. We have so far used the 2778 galaxy data cubes released in the 5th MaNGA Product Launch (MPL-5), obtained between 2014 and 2016. The cubes have been processed using the version 2.0.1 of the MaNGA Data Reduction Pipeline \citep{law16}. This set of datacubes contains 62 AGN hosts, selected using optical emission line diagnostics as described in Sect.~\ref{sec:agns}, comprising 45 low-luminosity ($L_{\rm[OIII]}<3.8\times 10^{40}$\,erg\,s$^{-1}$) AGN 
and 17 higher luminosity ones from both the baseline MaNGA sample and the Ancillary AGN sample\footnote{As the baseline MaNGA sample contains limited dynamic range in AGN luminosity, reaching only $L(\rm{[OIII]}) \approx 10^{40}$ erg~s$^{-1}$ ($L_{\rm bol} \approx 10^{43}$~erg~s$^{-1}$), it is not well suited to explore the relationship between the AGN power and both the outflows and stellar populations of the host galaxy.  In order to address this issue, an auxiliary program to cover a wider range of AGN luminosities was proposed by a group led by PI Jenny Greene. This successful proposal will lead to the observation of an additional $\approx$ 150 AGN with luminosities up to $L_{\rm bol} \approx 10^{45}$~erg~s$^{-1}$.}

Our goal with this first paper is to broadly characterize the properties of this initial AGN sample and select a control sample of inactive galaxies matched to the AGN hosts in absolute magnitude, galaxy mass, redshift, galaxy type and inclination. The definition of a control sample is essential in order to investigate the effects of AGN feeding and feedback on the host galaxy properties, and in order to do this, it is necessary to make sure that eventual differences are not related to properties such as Hubble type or galaxy mass. We have found so far 62 AGN and 109 control sample galaxies, but the criteria introduced in the present paper will be used to increase our sample to at least 300 AGN (estimated to be observed) and corresponding control sample galaxies by the conclusion of the MaNGA survey.

Besides presenting and characterizing the AGN and control samples, we investigate the similarities and differences between the stellar populations of the two samples. This study was performed in two steps: in the present paper we compare the stellar population of the AGN hosts and control galaxies using stellar population synthesis of the nuclear spectra over the inner 3 arcsec from the SDSS-III survey \citep{gunn06, eisenstein11, smee13}. In a forthcoming paper (Mallmann et al., in preparation) we will use the MaNGA datacubes to compare the resolved stellar population properties up to 1.5$r_e$. 

This paper is organized as follows. In Section 2 we present and describe the AGN and control samples; in Section 3 we  compare the properties of the two samples; in Section 4 we describe the stellar population synthesis method; in Section 5 we discuss the results and in Section 6 we present our conclusions. We have assumed a cosmology with $H_0 = 70\,\mbox{km\,s$^{-1}$\,Mpc$^{-1}$}$, $\Omega_m=0.3$ and $\Omega_V=0.7$.

\section{Sample selection and characterization}

\subsection{Galaxies hosting an active nucleus}
\label{sec:agns}

AGN produce a spectral energy distribution consistently
harder than massive main sequence stars. A common tool for identifying
the origin of the ionizing photons in optical emission-line galaxies
(active nuclei, starbursts or transition objects) is the BPT diagram
\citep{bpt81}, which is based on line ratios between high and low
ionization potential species. But one weakness of the BPT diagram is its inability to discriminate between a genuine low-ionization AGN and emission-line
galaxies whose ionizing photons are produced in the atmospheres of evolved
low-mass stars (the so-called post-AGB stars). In order to circumvent this limitation, \citet{cid10} have
introduced a new diagnostic diagram which makes use of the equivalent width of H$\alpha$ (EWH$\alpha$)
-- the so-called WHAN diagram. These authors have shown that galaxies that have EW$_{H\alpha}$ smaller than 3\AA{} are not ionized by AGN but instead are ionized by post-AGB stars, and have been dubbed ``LIERS". 

In order to identify hosts of ``true" AGN in the MaNGA MPL-5 sample, we have made use of the
BPT and the WHAN diagrams simultaneously. We have cross-matched all galaxy
data cubes observed in MPL-5 with the SDSS-III spectroscopic data from DR12 \citep{Alam+15}.
We have obtained line fluxes and
equivalent widths of H$\beta$, H$\alpha$, [OIII]$\lambda 5007$ and [NII]$\lambda 6584$,
measured in the SDSS-III integrated nuclear spectrum, from \citet{thomas+13}. The criterion we have applied to identify
galaxies hosting AGN was that it must be located,
within the uncertainties, simultaneously in the
Seyfert or LINER region of both the BPT and the WHAN diagrams. This
criterion elliminates from the sample potential LIERs, and also
transition objects. Only 62 galaxies have fulfilled this criterion.

Figures~\ref{bptw1} and \ref{bptw2} present the BPT and WHAN diagrams
for all MPL-5 emission line galaxies and indicate the location of galaxies
we have considered as AGN hosts following the criteria above.
Table~\ref{tableagns} lists relevant parameters of the AGN host galaxies. Besides identifications and coordinates, redshifts, absolute $r$ magnitudes and stellar masses, we also list the galaxy classification in Galaxy Zoo \citep[GZ1,][]{lintott11}, the $r$-band concentration index $C$  and asymmetry parameter $A$ (described below) and the [OIII]$\lambda 5007$ luminosity derived from SDSS-III data \citep{thomas+13}.

\begin{figure*}
\begin{center}
\begin{tabular}{ccc}
AGN & Control object 1 & Control object 2 \\
1-558912 & 1-71481 & 1-72928\\
\includegraphics[height=5cm]{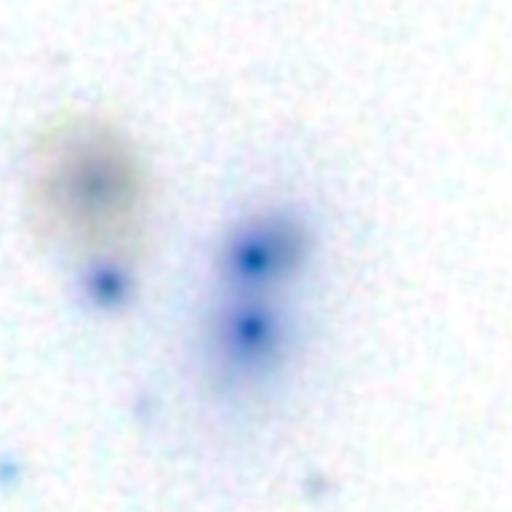} & \includegraphics[height=5cm]{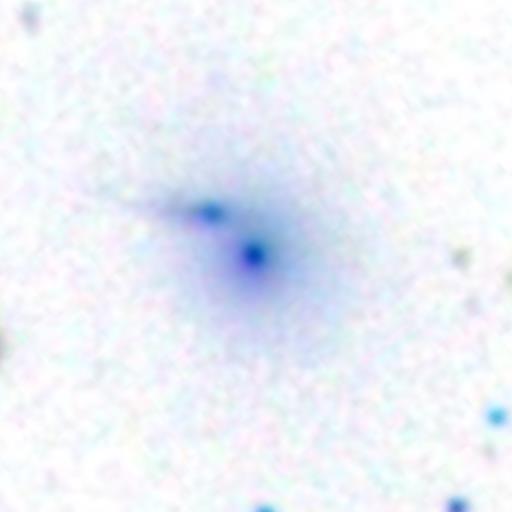} & \includegraphics[height=5cm]{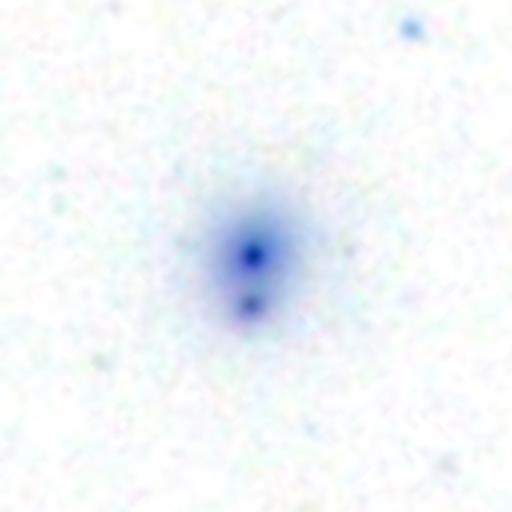} \\
1-269632 & 1-210700 & 1-378795\\
\includegraphics[height=5cm]{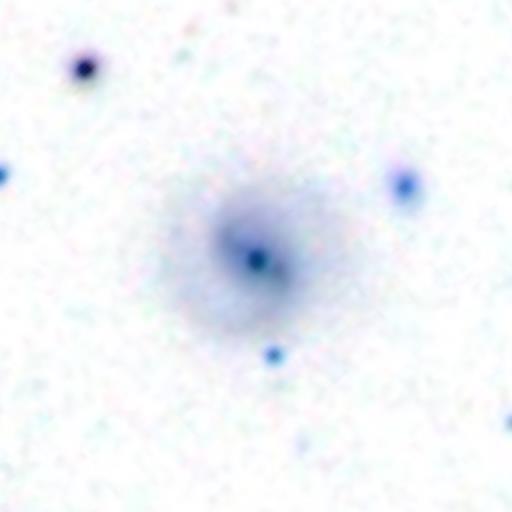} & \includegraphics[height=5cm]{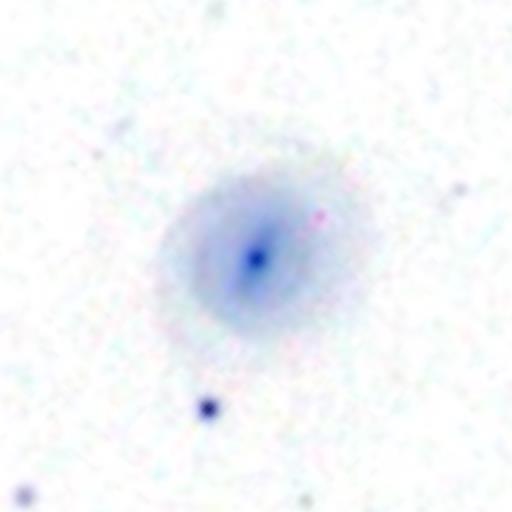} & \includegraphics[height=5cm]{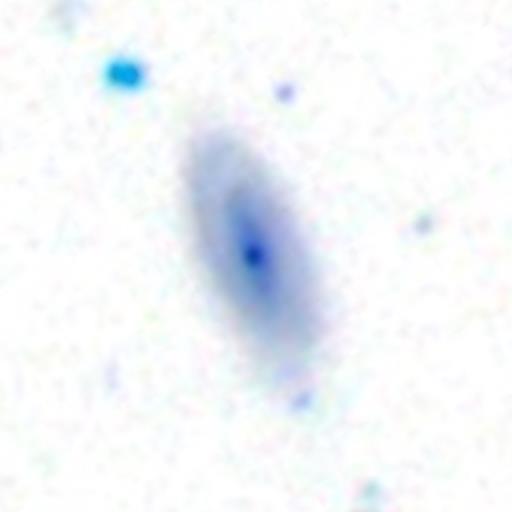} \\
1-258599 & 1-93876 & 1-166691\\
\includegraphics[height=5cm]{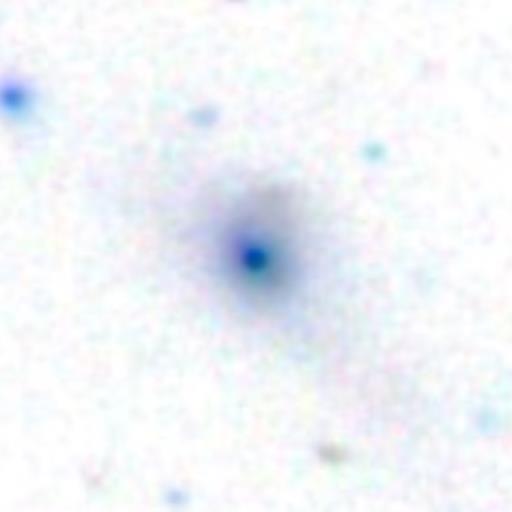} & \includegraphics[height=5cm]{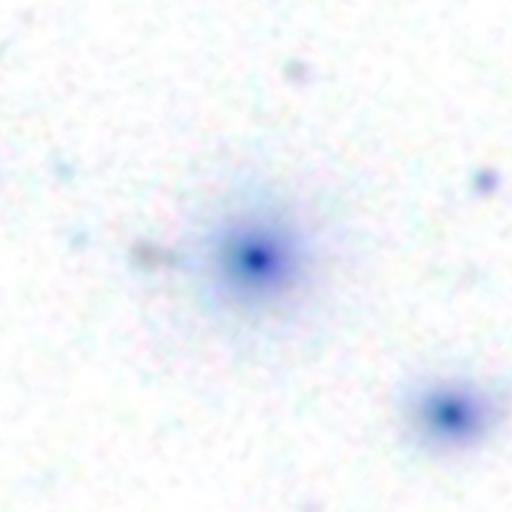} & \includegraphics[height=5cm]{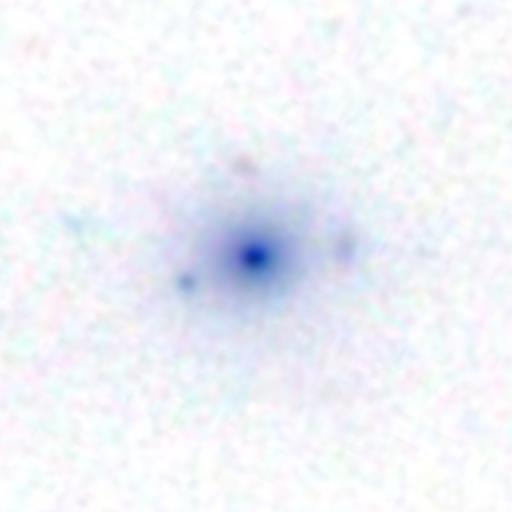} \\
1-72322 & 1-121717 & 1-43721\\
\includegraphics[height=5cm]{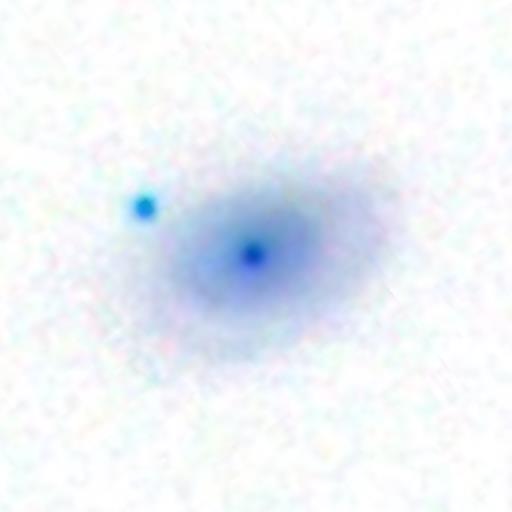} & \includegraphics[height=5cm]{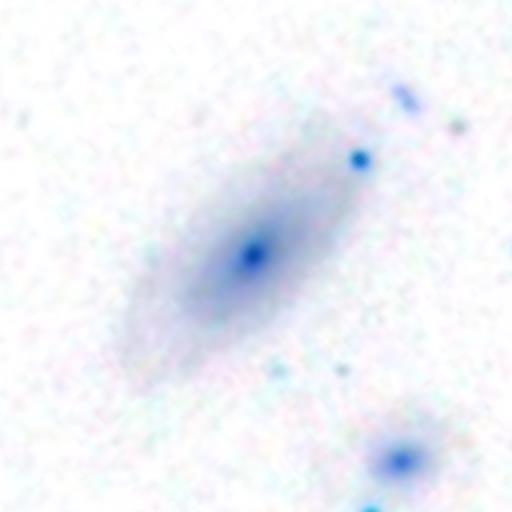} & \includegraphics[height=5cm]{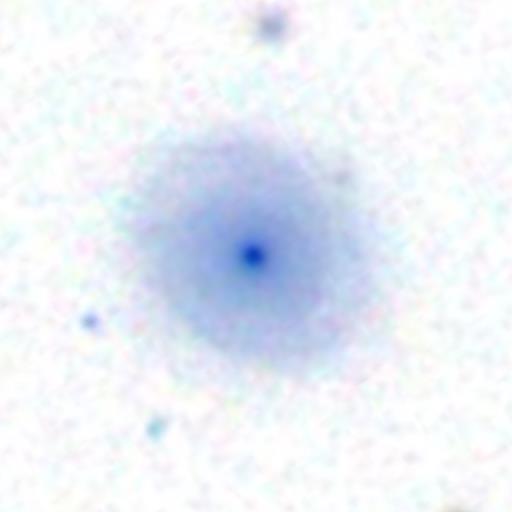} \\
\end{tabular}
\end{center}
\caption{SDSS-III $ugriz$ negative images of the four strongest AGN from our sample (left) and their control sample equivalents (center and right columns). The mangaID of each host is indicated above its respective panel. Each chart is 51 arcsec on a side; north is up, east to the left.}
\label{agnimages}
\end{figure*}

In order to assess quantitatively the morphological variety of the AGN and control samples, we have derived the concentration index $C$ and asymmetry index $A$ of the sample galaxies using the publicly available code \textsc{PyCA} \citep{menanteau05}. This calculation was made using the $r$-band images in order to maximize the galaxy signal-to-noise ratio. We have also obtained for all galaxies of the sample the 
morphological classifications from GZ1 \citep{lintott11}. From the GZ1 classification probabilities, we have separated the galaxies in four large classes: ellipticals (E), spirals (S), merging (M) and intermediate elliptical/spiral (E/S); this last class corresponds to galaxies whose GZ1 elliptical and spiral probabilities are rigorously the same and exceed 50\% when combined.

Most of the 62 AGN in our sample have luminosities $L(\rm{[OIII]})< 3.8\times 10^{40}$\,erg\,s$^{-1}$, with only 17 having larger luminosities. We use this  threshold -- postulated by \citet{kauffmann03} -- to characterize these 17 AGN as ``strong AGN".  SDSS-III \textit{ugriz} negative images \citep{fukugita96} of the four most luminous of these strong AGN are shown in Figure~\ref{agnimages}; the remainder are shown in the Appendix. 

Inspection of the SDSS-III spectra showed that most of the strong AGN in our sample present coronal lines:
at least one of the lines [NeV]$\lambda$3425\AA, [FeVII]$\lambda$3760\AA\ or [FeX]$\lambda$6374\AA\ are detectable above $3\sigma$.
A small number of objects among the most [OIII]-luminous ``weak AGN'' also present some of these
lines. Coronal lines are
thought to be produced by photoionization from the AGN \citep{mazzalay+10,ardila+11} and to be associated with outflows \citep{mullersanchez+11}. Being insensitive to star formation as opposed to the [OIII]$\lambda$5007\AA, the presence of coronal lines reinforces the AGN nature of the 17 most [OIII]-luminous objects in our sample, and suggests a transition between high- and low-ionization AGN at $L(\rm{[OIII]}) =3.8\times 10^{40}$\,erg\,s$^{-1}$, in concordance with \citet{kauffmann03}.

We note that our AGN selection is based on single-fibre SDSS-III observations which limits us to sources with AGN-photoionization signatures within 3 arcsec of the galactic center. \citet{Wylezalek17} has recently developed an AGN selection algorithm taking full advantage of the spatial dimension of AGN ionization signatures provided through the MaNGA data. \citet{Wylezalek17} show that about a third to a half of the MaNGA-selected AGN candidates would not have been selected based on the SDSS-III single-fiber observations since AGN ionization signatures are only prevalent beyond the 3 arcsec coverage of the single-fiber spectra. Reasons for this can be manyfold (off-nuclear AGN, star formation signatures dominate in the center due to a nuclear starburst, recently turned-off AGN) and are currently under investigation by Wylezalek et al. in prep.

In this paper, we focus on the classical AGN with nuclear photo-ionization signatures. In a forthcoming paper, we will also investigate the nature of the stellar populations in the ``unusual'' off-nuclear MaNGA-selected AGN candidates.

\begin{table*}
\caption{Parameters of AGN in MaNGA-MPL5. (1) galaxy identification in the MaNGA survey; (2) MaNGA plate-IFU identification of the observation; (3)-(4): RA/DEC (2000) in degrees; (5) spectroscopic redshift from SDSS-III; (6): integrated
absolute $r$-band magnitude from SDSS-III; (7): stellar mass in units of $M_\odot$; errors associated to the stellar masses of galaxies in our sample are typically under 0.03\,dex \citep{Conroy+09}; (8) elliptical/spiral/merging classification from Galaxy Zoo I; (9)-(10): $r$-band concentration and asymmetry derived with PyCA; (11) [OIII] luminosity in units of $10^{40}$\,erg\,s$^{-1}$.}
\begin{tabular}{ccccccccccccc}
\hline
mangaID & RA  & DEC & $z$ & $M_r$ & log $M^\star/M_\odot$ & GZ1$_c$ & $C$ & $A$  & $L(\rm{[OIII]})$\\
 (1)    & (2)      & (3) & (4) & (5) & (6)   & (7)                   & (8)     & (9) & (10) \\
\hline
1-558912 & 166.129410 & 42.624554 & 0.1261 & -20.46 & 11.25 & -- & 0.37 & 0.12 & 56.82$\pm$1.25\\
1-269632 & 247.560974 & 26.206474 & 0.1315 & -21.78 & 11.62 & S & 0.47 & 0.05 & 30.08$\pm$1.69\\
1-258599 & 186.181000 & 44.410770 & 0.1256 & -21.24 & 11.68 & E & 0.50 & 0.11 & 20.95$\pm$0.67\\
1-72322 & 121.014198 & 40.802612 & 0.1262 & -21.81 & 12.05 & S & 0.34 & 0.08 & 20.66$\pm$0.43\\
1-121532 & 118.091110 & 34.326569 & 0.1400 & -20.51 & 11.34 & E & 0.33 & 0.05 & 11.68$\pm$0.96\smallskip\\
1-209980 & 240.470871 & 45.351940 & 0.0420 & -19.70 & 10.79 & S & 0.57 & 0.04 & 11.01$\pm$0.17\\
1-44379 & 120.700706 & 45.034554 & 0.0389 & -19.89 & 10.97 & S & 0.24 & 0.06 & 8.94$\pm$0.14\\
1-149211 & 168.947800 & 50.401634 & 0.0473 & -18.27 & 10.16 & S & 0.29 & 0.03 & 7.88$\pm$0.14\\
1-173958 & 167.306015 & 49.519432 & 0.0724 & -20.53 & 11.31 & S & 0.33 & 0.06 & 6.79$\pm$0.30\\
1-338922 & 114.775749 & 44.402767 & 0.1345 & -20.27 & 11.13 & M & 0.44 & 0.03 & 6.77$\pm$0.90\smallskip\\
1-279147 & 168.957733 & 46.319565 & 0.0533 & -19.51 & 10.66 & S & 0.45 & 0.03 & 6.77$\pm$0.20\\
1-460812 & 127.170799 & 17.581400 & 0.0665 & -19.81 & 11.44 & -- & 0.38 & 0.05 & 6.46$\pm$0.31\\
1-92866 & 243.581818 & 50.465611 & 0.0603 & -20.56 & 11.69 & E & 0.49 & 0.05 & 6.12$\pm$0.30\\
1-94784 & 249.318420 & 44.418228 & 0.0314 & -20.06 & 10.85 & S & 0.42 & 0.03 & 5.96$\pm$0.12\\
1-44303 & 119.182152 & 44.856709 & 0.0499 & -19.72 & 10.62 & S & 0.29 & 0.10 & 5.56$\pm$0.12\smallskip\\
1-339094 & 117.472420 & 45.248482 & 0.0313 & -19.02 & 10.52 & E & 0.36 & 0.03 & 5.29$\pm$0.09\\
1-137883 & 137.874756 & 45.468319 & 0.0268 & -18.06 & 10.77 & E/S & 0.41 & 0.01 & 3.87$\pm$0.12\\
1-48116 & 132.653992 & 57.359669 & 0.0261 & -19.18 & 10.60 & S & 0.31 & 0.06 & 3.79$\pm$0.08\\
1-256446 & 166.509872 & 43.173473 & 0.0584 & -19.40 & 11.14 & E & 0.49 & 0.05 & 3.74$\pm$0.15\\
1-95585 & 255.029877 & 37.839500 & 0.0633 & -20.88 & 11.24 & S & 0.27 & 0.08 & 3.58$\pm$0.16\smallskip\\
1-135641 & 249.557312 & 40.146820 & 0.0304 & -19.03 & 11.19 & S & 0.28 & 0.08 & 3.52$\pm$0.09\\
1-259142 & 193.703995 & 44.155567 & 0.0543 & -20.75 & 11.29 & S & 0.39 & 0.06 & 3.47$\pm$0.20\\
1-109056 & 39.446587 & 0.405085 & 0.0473 & -19.27 & 10.57 & -- & 0.32 & 0.05 & 3.24$\pm$0.08\\
1-24148 & 258.827423 & 57.658772 & 0.0282 & -18.51 & 10.56 & S & 0.31 & 0.04 & 3.17$\pm$0.05\\
1-166919 & 146.709106 & 43.423843 & 0.0722 & -20.85 & 11.28 & S & 0.37 & 0.06 & 2.64$\pm$0.25\smallskip\\
1-248389 & 240.658051 & 41.293427 & 0.0348 & -19.36 & 10.57 & S & 0.49 & 0.12 & 2.55$\pm$0.09\\
1-321739 & 226.431656 & 44.404903 & 0.0283 & -18.91 & 11.12 & S & 0.40 & 0.14 & 2.24$\pm$0.10\\
1-234618 & 202.128433 & 47.714039 & 0.0608 & -19.64 & 11.37 & S & 0.31 & 0.09 & 2.23$\pm$0.23\\
1-229010 & 57.243038 & -1.144831 & 0.0407 & -20.51 & 11.46 & -- & 0.41 & 0.03 & 2.11$\pm$0.09\\
1-211311 & 248.426392 & 39.185120 & 0.0298 & -19.04 & 10.44 & E/S & 0.43 & 0.02 & 1.99$\pm$0.06\smallskip\\
1-373161 & 222.810074 & 30.692245 & 0.0547 & -21.30 & 11.60 & E & 0.43 & 0.00 & 1.87$\pm$0.11\\
1-210646 & 245.157181 & 41.466873 & 0.0606 & -20.38 & 10.98 & S & 0.18 & 0.05 & 1.80$\pm$0.10\\
1-351790 & 121.147926 & 50.708557 & 0.0227 & -18.09 & 9.92 & E & 0.39 & 0.02 & 1.72$\pm$0.03\\
1-163831 & 118.627846 & 25.815987 & 0.0631 & -20.84 & 11.26 & S & 0.27 & 0.05 & 1.67$\pm$0.13\\
1-22301 & 253.405563 & 63.031269 & 0.1052 & -21.19 & 11.18 & S & 0.29 & 0.08 & 1.67$\pm$0.23\smallskip\\
1-248420 & 241.823395 & 41.403603 & 0.0346 & -19.71 & 10.90 & S & 0.21 & 0.07 & 1.66$\pm$0.06\\
1-23979 & 258.158752 & 57.322422 & 0.0266 & -18.27 & 10.42 & E & 0.44 & 0.06 & 1.60$\pm$0.05\\
1-542318 & 245.248306 & 49.001778 & 0.0582 & -19.75 & 10.91 & E & 0.34 & 0.01 & 1.58$\pm$0.07\\
1-95092 & 250.846420 & 39.806461 & 0.0302 & -19.95 & 11.20 & E & 0.47 & 0.04 & 1.54$\pm$0.07\\
1-279676 & 173.981888 & 48.021458 & 0.0587 & -19.40 & 10.81 & -- & 0.32 & 0.02 & 1.52$\pm$0.14\smallskip\\
1-201561 & 118.053215 & 28.772579 & 0.0637 & -19.73 & 10.88 & S & 0.30 & 0.07 & 1.37$\pm$0.15\\
1-198182 & 224.749649 & 48.409855 & 0.0359 & -20.22 & 11.09 & E & 0.49 & 0.01 & 1.34$\pm$0.11\\
1-96075 & 253.946381 & 39.310535 & 0.0631 & -21.12 & 11.35 & S & 0.29 & 0.07 & 1.26$\pm$0.13\\
1-519742 & 206.612457 & 22.076742 & 0.0276 & -17.62 & 9.64 & S & 0.23 & 0.04 & 1.19$\pm$0.03\\
1-491229 & 172.607544 & 22.216530 & 0.0393 & -20.25 & 11.12 & E & 0.51 & 0.02 & 1.14$\pm$0.11\smallskip\\
1-604761 & 113.472275 & 37.025906 & 0.0618 & -20.92 & 11.34 & S & 0.26 & 0.12 & 1.00$\pm$0.13\\
1-25725 & 262.996735 & 59.971638 & 0.0291 & -18.30 & 10.55 & E & 0.44 & 0.04 & 0.92$\pm$0.05\\
1-94604 & 251.335938 & 42.757790 & 0.0493 & -19.44 & 10.52 & S & 0.37 & 0.01 & 0.86$\pm$0.07\\
1-37036 & 41.699909 & 0.421577 & 0.0283 & -19.02 & 10.66 & E & 0.40 & 0.09 & 0.84$\pm$0.06\\
1-167688 & 155.885559 & 46.057755 & 0.0258 & -17.86 & 9.75 & E & 0.52 & 0.04 & 0.84$\pm$0.02\smallskip\\
1-279666 & 173.911240 & 47.515518 & 0.0455 & -18.83 & 10.42 & E & 0.31 & 0.02 & 0.84$\pm$0.07\\
1-339163 & 116.280205 & 46.072422 & 0.0312 & -20.02 & 10.97 & S & 0.30 & 0.10 & 0.82$\pm$0.07\\
1-258774 & 186.400864 & 45.083858 & 0.0384 & -19.60 & 10.77 & -- & 0.55 & 0.03 & 0.77$\pm$0.10\\
1-198153 & 224.289078 & 48.633968 & 0.0354 & -19.83 & 11.00 & S & 0.27 & 0.07 & 0.76$\pm$0.08\\
1-91016 & 234.810974 & 56.670856 & 0.0463 & -18.60 & 10.56 & S & 0.27 & 0.06 & 0.76$\pm$0.09\smallskip\\
1-279073 & 170.588150 & 46.430504 & 0.0323 & -19.53 & 10.79 & E & 0.51 & 0.01 & 0.63$\pm$0.06\\
1-135044 & 247.907990 & 41.493645 & 0.0303 & -19.76 & 10.65 & S & 0.31 & 0.05 & 0.61$\pm$0.04\\
\hline
\end{tabular}
\label{tableagns}
\end{table*}

\begin{table*}
\contcaption{}
\begin{tabular}{ccccccccccccc}
\hline
mangaID & RA  & DEC & $z$ & $M_r$ & log $M^\star/M_\odot$ & GZ1$_c$ & $C$ & $A$  & $L(\rm{[OIII]})$\\
 (1)    & (2)      & (3) & (4) & (5) & (6)   & (7)                   & (8)     & (9) & (10) \\
\hline
1-148068 & 156.805679 & 48.244793 & 0.0610 & -20.72 & 11.41 & S & 0.22 & 0.04 & 0.45$\pm$0.15\\
1-277552 & 167.034561 & 45.984623 & 0.0362 & -19.72 & 10.83 & S & 0.21 & 0.15 & 0.44$\pm$0.05\\
1-217050 & 136.719986 & 41.408253 & 0.0274 & -19.66 & 10.93 & E & 0.47 & 0.02 & 0.43$\pm$0.03\smallskip\\
1-25554 & 262.486053 & 58.397408 & 0.0268 & -19.27 & 10.52 & S & 0.36 & 0.04 & 0.24$\pm$0.03\\
1-135285 & 247.216949 & 42.812012 & 0.0316 & -19.66 & 10.78 & -- & 0.32 & 0.05 & 0.20$\pm$0.04\\
\hline
\end{tabular}
\end{table*}

\subsection{Control sample}
\label{sec:cs}

We have extracted from MPL-5 a control sample of galaxies with properties matched to those of
the AGN hosts, except that their nuclei are quiescent. The control sample was built as follows:
\begin{enumerate}
\item First, we selected from MPL-5 all galaxies, presenting or not detectable emission lines, whose ionizing source
is not an AGN (as discussed in sec. 2.1).
We therefore consider a galaxy as a potential control sample candidate if it
fulfills any one of these four conditions: (a)
it is located, within the uncertainties, in the starforming region
of the BPT diagram; (b) it is located, within the uncertainties, in the transition or LINER region of the BPT, but the WHAN diagram discards ionization from an AGN based on
the value of EWH$\alpha$; (c) it presents a very large uncertainty in
$\log$[OIII]/H$\beta$, not allowing for a secure classification in the BPT diagram,
but the WHAN diagram discards ionization from an AGN based on
the value of EWH$\alpha$; (d) the galaxy does not present emission lines. These conditions make sure that galaxies
with prominent emission lines are dominated by star formation, not by an active nucleus;
the remaining objects are galaxies whose emission lines can be completely
accounted for by hot evolved low-mass stars. Galaxies which fulfill
any one of these conditions are considered inactive.
\item We have then selected two control galaxies for each AGN, matching then according to the redshift $z$ and total stellar mass $M_\star$. A preliminary list of control sample candidates was drawn for each galaxy in the AGN sample by exploring the full $(z,M_\star)$ parameter space
for all inactive galaxies in MPL-5. Galaxies for which both $z$ and $M_\star$ did not differ by more than $30\%$ from those of the AGN sample values were visually inspected, allowing for a large number of control objects to be selected while keeping low the dispersion of values in the parameter space. The number of control sample candidates selected at this point for each galaxy was typically $\approx$ 50 (with a larger/smaller number of candidates for galaxies at low/high redshift).
\item The images of the control candidates for each galaxy in the AGN sample were then visually inspected to select control galaxies with similar morphologies and axial ratios to those of the AGN host galaxies. For late-type galaxies, we have tried to reproduce both the structure and development of the arms and the relative bulge size, as well as the line-of-sight orientation and the presence of dust lanes. The priority in the morphological analysis was given to structures closer to the nucleus; i.e., we have assigned more weight to the presence of bars and rings than to the morphology of the arms in the outer galaxy regions. During this step the number of potentially good candidates, i.e. galaxies which are morphologically comparable to the respective AGN host, decreased in many cases to only two galaxies. In order to produce the best statistical comparison with the AGN sample properties, keeping an homogeneous representation of all AGN hosts in the control sample, only the two galaxies which best matched the morphology of the AGN host, according to the above criteria, were chosen for each AGN in the sample.
\end{enumerate}

\begin{figure*}
\centering
\subfloat[]{\label{bptw1}\includegraphics[width=\columnwidth]{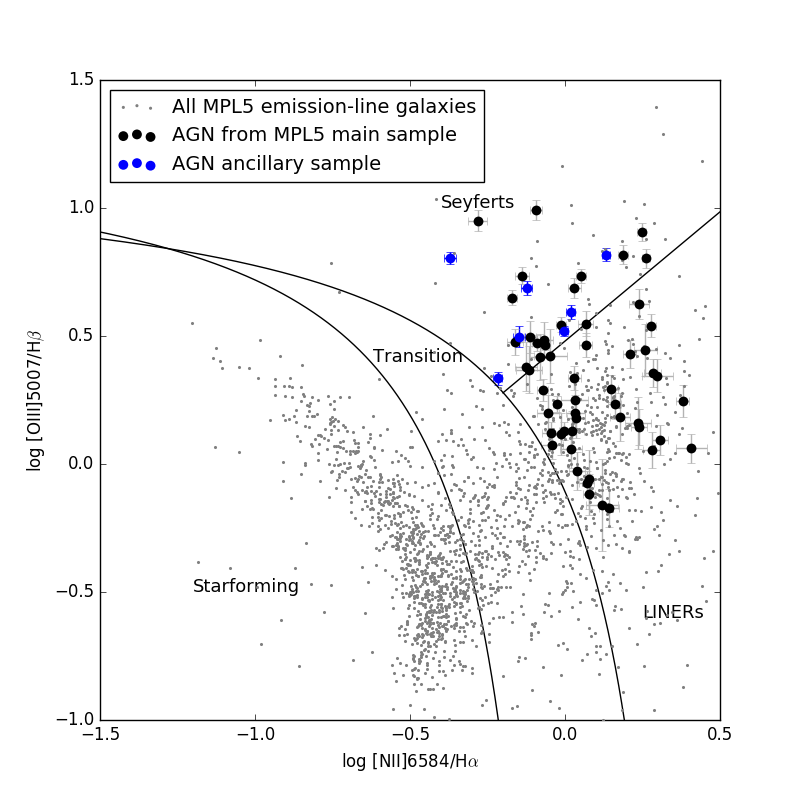}}
\subfloat[]{\label{bptw2}\includegraphics[width=\columnwidth]{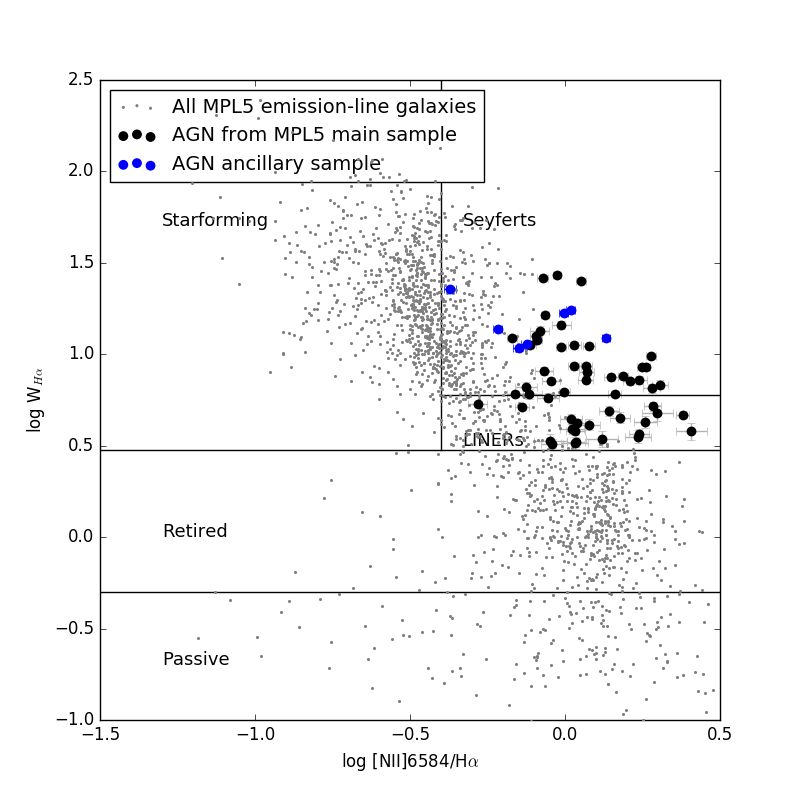}}\\
\subfloat[]{\label{bptw3}\includegraphics[width=\columnwidth]{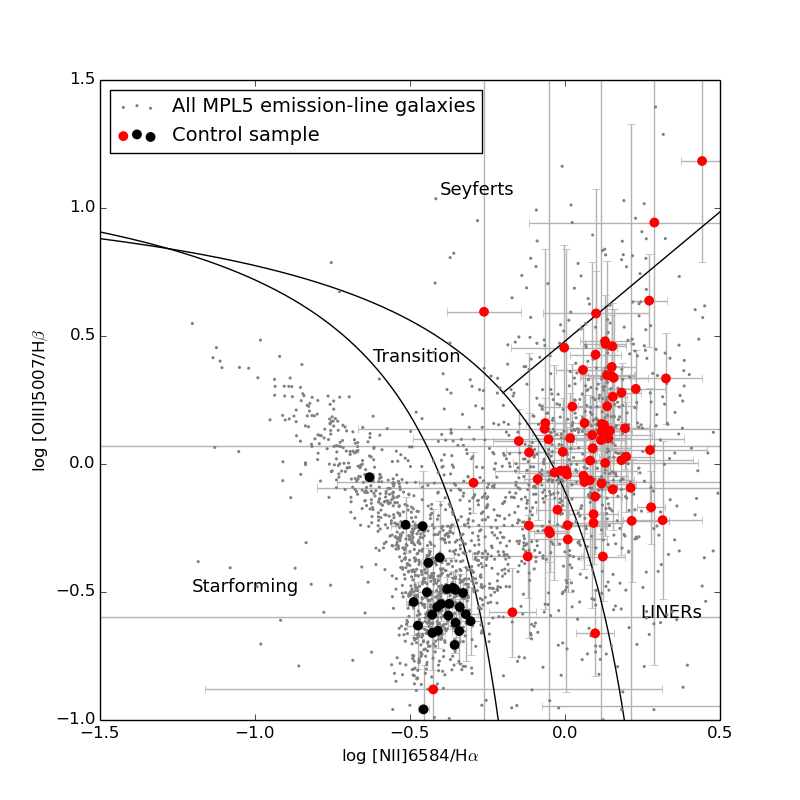}}
\subfloat[]{\label{bptw4}\includegraphics[width=\columnwidth]{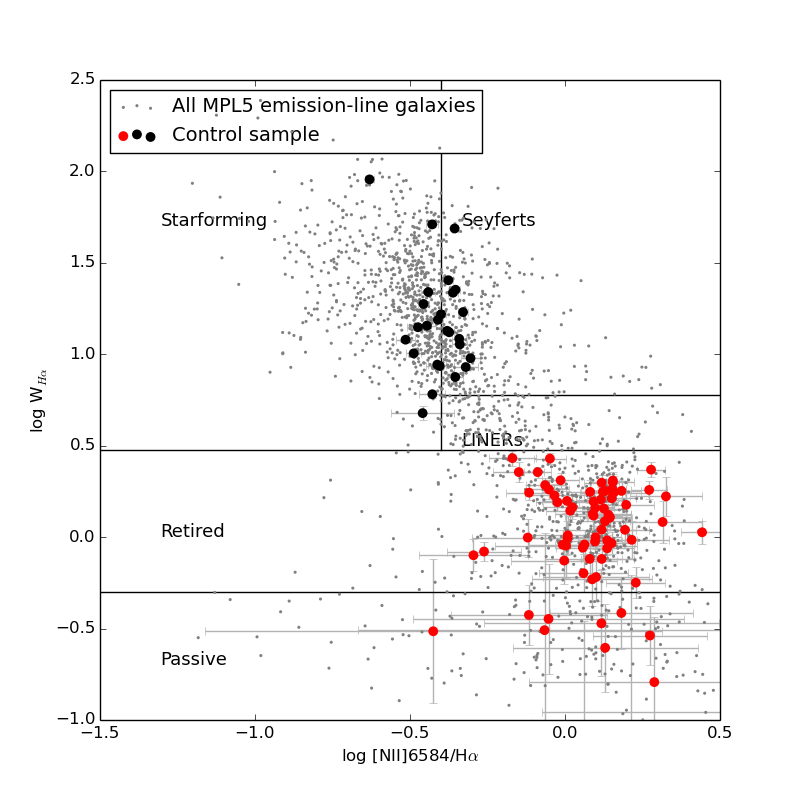}}
\caption{BPT and WHAM diagrams for our sample of confirmed AGN (top panels) and control sample (bottom panels), using the fluxes from \citet{thomas+13}. We only include objects where the emission lines are detected at the 3-$\sigma$ level.
The continuous lines separating Seyferts, transition objects, LINERs and star-forming galaxies are from \citet{kauffmann03}, \citet{kewley01} and \citet{cid10}. Grey dots indicate the position of all emission-line galaxies in MPL-5. In the top panels, we indicate as black (blue) circles the confirmed AGN drawn from the MaNGA MPL-5 main sample (AGN Ancillary Sample). In the bottom diagrams, star-forming (retired/passive) control sample objects are represented by black (red) circles. 
Six objects from the control sample do not present detectable emission lines and are not shown in the diagrams.}
\end{figure*}

We have thus produced 124 pairs of AGN and control partners. Twelve of the control sample objects were selected as ``control partners'' of two or more AGN hosts simultaneously; therefore, our control sample comprises 109 non-active galaxies, reproducing the distribution of redshift and stellar mass of the AGN sample. Figures~\ref{bptw3}
and \ref{bptw4} present the BPT and WHAN diagrams for the control sample, superimposed
on the distribution of the full MPL-5 sample. Six control sample galaxies did not present any emission lines, and thus do not apper in the figures. We can see
a bimodal distribution with a clear separation between starforming galaxies
and LIERs. Comparing with the diagrams shown in Figures \ref{bptw1} and \ref{bptw2},
we can see that AGN and non-AGN have markedly distinct distributions.

\begin{table*}
\caption{Control sample parameters. (1) identification of the AGN host associated to the control galaxy; (2)-(12) same as (1)-(11) of Table~\ref{tableagns}. Twelve control sample objects
have been paired to more than one AGN host and appear more than once in the table.}
\begin{tabular}{ccccccccccccc}
\hline
AGN mangaID & mangaID & RA  & DEC & $z$ & $M_r$ & log $M^\star/M_\odot$ & GZ1$_c$ & $C$ & $A$  & $L(\rm{[OIII]})$\\
 (1)    & (2)      & (3) & (4) & (5) & (6)   & (7)                   & (8)     & (9) & (10) & (11) \\
\hline
1-558912 & 1-71481 & 117.456001 & 34.883911 & 0.1312 & -20.95 & 11.70 & E & 0.47 & 0.02 & 0.10$\pm$0.20\\
         & 1-72928 & 127.256485 & 45.016773 & 0.1270 & -20.62 & 11.52 & E & 0.40 & 0.21 & 0.09$\pm$0.23\\
1-269632 & 1-210700 & 248.140564 & 39.131020 & 0.1303 & -20.96 & 11.67 & S & 0.36 & 0.03 & 1.55$\pm$0.44\\
         & 1-378795 & 118.925613 & 50.172771 & 0.0967 & -20.77 & 11.35 & S & 0.32 & 0.03 & 0.72$\pm$0.31\smallskip\\
1-258599 & 1-93876 & 246.942947 & 44.177521 & 0.1394 & -20.75 & 11.50 & E & 0.44 & 0.01 & 0.46$\pm$0.36\\
         & 1-166691 & 146.047348 & 42.900040 & 0.1052 & -20.50 & 11.36 & E & 0.51 & 0.04 & 0.09$\pm$0.49\\
1-72322 & 1-121717 & 118.803429 & 35.596798 & 0.1098 & -21.11 & 11.61 & S & 0.39 & 0.12 & 1.40$\pm$0.57\\
         & 1-43721 & 116.967567 & 43.383499 & 0.1114 & -21.41 & 11.86 & S & 0.32 & 0.01 & 1.91$\pm$0.52\smallskip\\
1-121532 & 1-218427 & 124.342316 & 27.796206 & 0.1496 & -21.30 & 11.47 & E & 0.47 & 0.04 & 0.72$\pm$0.62\\
         & 1-177493 & 257.085754 & 31.746916 & 0.1081 & -20.90 & 11.30 & E & 0.38 & 0.06 & 2.29$\pm$0.28\\
1-209980 & 1-295095 & 248.348663 & 24.776577 & 0.0410 & -18.40 & 10.14 & E & 0.35 & 0.05 & 0.15$\pm$0.03\\
         & 1-92626 & 241.799545 & 48.572563 & 0.0434 & -20.04 & 11.04 & S & 0.36 & 0.03 & 0.76$\pm$0.07\smallskip\\
1-44379 & 1-211082 & 247.620041 & 39.626045 & 0.0304 & -19.72 & 11.07 & E & 0.31 & 0.06 & 0.19$\pm$0.04\\
         & 1-135371 & 250.156235 & 39.221634 & 0.0352 & -19.20 & 10.76 & S & 0.28 & 0.11 & 0.25$\pm$0.07\\
1-149211 & 1-377321 & 110.556152 & 42.183643 & 0.0444 & -19.02 & 9.89 & S & 0.31 & 0.03 & 4.53$\pm$0.13\\
         & 1-491233 & 172.563995 & 22.992010 & 0.0332 & -18.39 & 10.59 & S & 0.29 & 0.06 & 0.25$\pm$0.03\smallskip\\
1-173958 & 1-247456 & 232.823196 & 45.416538 & 0.0705 & -20.05 & 10.83 & -- & 0.40 & 0.02 & 0.57$\pm$0.16\\
         & 1-24246 & 264.840790 & 56.567070 & 0.0818 & -19.91 & 10.57 & S & 0.75 & 0.36 & 0.11$\pm$0.06\\
1-338922 & 1-286804 & 211.904861 & 44.482269 & 0.1429 & -20.03 & 10.50 & M & 0.44 & 0.32 & 2.23$\pm$0.43\\
         & 1-109493 & 56.425140 & -0.378460 & 0.1093 & -20.46 & 11.26 & -- & 0.49 & -0.01 & 0.15$\pm$0.18\smallskip\\
1-279147 & 1-283246 & 191.078873 & 46.407131 & 0.0496 & -19.17 & 10.55 & S & 0.47 & 0.04 & 0.23$\pm$0.06\\
         & 1-351538 & 119.145126 & 47.563850 & 0.0692 & -19.67 & 11.00 & S & 0.35 & 0.08 & 0.46$\pm$0.13\\
1-460812 & 1-270160 & 248.274612 & 26.211815 & 0.0660 & -20.37 & 11.46 & S & 0.50 & 0.02 & 0.70$\pm$0.39\\
         & 1-258455 & 183.612198 & 45.195454 & 0.0653 & -20.02 & 11.03 & E & 0.40 & 0.03 & 0.49$\pm$0.14\smallskip\\
1-92866 & 1-94514 & 248.241180 & 42.524670 & 0.0614 & -20.60 & 11.17 & E & 0.51 & 0.00 & --\\
         & 1-210614 & 244.501755 & 41.392189 & 0.0612 & -20.64 & 11.48 & E & 0.49 & 0.01 & 0.40$\pm$0.14\\
1-94784 & 1-211063 & 247.058411 & 40.313835 & 0.0331 & -19.87 & 10.79 & S & 0.33 & 0.09 & 0.20$\pm$0.04\\
         & 1-135502 & 247.764175 & 39.838505 & 0.0305 & -19.51 & 11.13 & S & 0.40 & 0.09 & 0.50$\pm$0.05\smallskip\\
1-44303 & 1-339028 & 116.097923 & 44.527740 & 0.0497 & -20.01 & 11.24 & S & 0.36 & 0.06 & 0.44$\pm$0.08\\
         & 1-379087 & 119.910118 & 51.792362 & 0.0534 & -19.60 & 11.02 & S & 0.38 & 0.10 & 0.72$\pm$0.13\\
1-339094 & 1-274646 & 158.017029 & 43.859268 & 0.0284 & -18.70 & 10.36 & E & 0.53 & 0.02 & 0.35$\pm$0.04\\
         & 1-24099 & 258.027618 & 57.504009 & 0.0282 & -18.67 & 10.34 & E & 0.44 & 0.01 & 0.06$\pm$0.03\smallskip\\
1-137883 & 1-178838 & 312.023621 & 0.068841 & 0.0247 & -17.54 & 10.46 & -- & 0.51 & 0.19 & 0.10$\pm$0.02\\
         & 1-36878 & 42.542126 & -0.867116 & 0.0232 & -18.88 & 10.77 & E & 0.45 & 0.07 & 0.28$\pm$0.04\\
1-48116 & 1-386452 & 136.228333 & 28.384314 & 0.0269 & -19.54 & 10.57 & S & 0.49 & 0.09 & 0.32$\pm$0.04\\
         & 1-24416 & 263.033173 & 56.878746 & 0.0281 & -19.16 & 10.66 & S & 0.37 & 0.03 & 0.22$\pm$0.03\smallskip\\
1-256446 & 1-322671 & 235.797028 & 39.238773 & 0.0637 & -19.77 & 10.82 & E & 0.49 & 0.04 & --\\
         & 1-256465 & 166.752243 & 43.089901 & 0.0575 & -19.70 & 10.79 & E & 0.50 & 0.01 & 0.59$\pm$0.11\\
1-95585 & 1-166947 & 147.335007 & 43.442989 & 0.0720 & -20.79 & 10.81 & S & 0.29 & 0.02 & 0.13$\pm$0.08\\
         & 1-210593 & 244.419754 & 41.899155 & 0.0605 & -19.76 & 10.90 & S & 0.36 & 0.06 & 0.43$\pm$0.14\smallskip\\
1-135641 & 1-635503 & 318.990448 & 9.543076 & 0.0293 & -19.37 & 10.91 & S & 0.22 & 0.10 & 0.15$\pm$0.06\\
         & 1-235398 & 213.149185 & 47.253059 & 0.0281 & -18.91 & 10.99 & S & 0.28 & 0.10 & 0.16$\pm$0.05\\
1-259142 & 1-55572 & 133.121307 & 56.112690 & 0.0454 & -20.11 & 11.03 & S & 0.40 & 0.06 & 0.12$\pm$0.04\\
         & 1-489649 & 171.954834 & 21.386103 & 0.0406 & -19.94 & 10.95 & S & 0.40 & 0.03 & 0.30$\pm$0.08\smallskip\\
1-109056 & 1-73005 & 125.402306 & 45.585476 & 0.0514 & -19.47 & 10.65 & S & 0.31 & 0.05 & 0.20$\pm$0.06\\
         & 1-43009 & 113.553879 & 39.076836 & 0.0510 & -19.41 & 10.43 & S & 0.26 & 0.03 & 0.12$\pm$0.04\\
1-24148 & 1-285031 & 198.701370 & 47.351547 & 0.0303 & -18.47 & 10.72 & S & 0.34 & 0.05 & 0.26$\pm$0.04\\
         & 1-236099 & 225.236221 & 41.566265 & 0.0205 & -17.36 & 9.91 & S & 0.33 & 0.04 & 0.07$\pm$0.01\smallskip\\
1-166919 & 12-129446 & 203.943542 & 26.101791 & 0.0670 & -20.57 & 11.32 & S & 0.34 & 0.03 & 0.28$\pm$0.09\\
         & 1-90849 & 237.582748 & 56.131981 & 0.0661 & -20.39 & 11.16 & E & 0.30 & 0.04 & 0.28$\pm$0.05\\
1-248389 & 1-94554 & 248.914688 & 42.461296 & 0.0318 & -18.96 & 10.57 & S & 0.55 & 0.07 & 0.22$\pm$0.04\\
         & 1-245774 & 214.863297 & 54.100300 & 0.0426 & -20.22 & 10.83 & S & 0.40 & 0.08 & 0.29$\pm$0.07\smallskip\\
1-321739 & 1-247417 & 233.319382 & 45.698528 & 0.0294 & -19.25 & 10.76 & S & 0.28 & 0.09 & 0.16$\pm$0.04\\
         & 1-633994 & 247.419952 & 40.686954 & 0.0305 & -18.27 & 11.04 & S & 0.39 & 0.11 & 0.36$\pm$0.09\\
1-234618 & 1-282144 & 184.592514 & 46.155350 & 0.0492 & -18.92 & 10.31 & S & 0.21 & 0.08 & 0.10$\pm$0.02\\
         & 1-339125 & 117.739944 & 45.989529 & 0.0534 & -18.97 & 11.17 & S & 0.35 & 0.05 & 0.45$\pm$0.23\smallskip\\
1-229010 & 1-210962 & 246.358719 & 39.870697 & 0.0290 & -20.49 & 11.09 & S & 0.47 & 0.07 & 0.35$\pm$0.06\\
         & 1-613211 & 167.861847 & 22.970764 & 0.0323 & -19.87 & 11.32 & E & 0.48 & 0.04 & 0.16$\pm$0.06\\
\hline
\end{tabular}
\label{tablecontrol}
\end{table*}

\begin{table*}
\contcaption{}
\begin{tabular}{ccccccccccccc}
\hline
AGN mangaID & mangaID & RA  & DEC & $z$ & $M_r$ & log $M^\star/M_\odot$ & GZ1$_c$ & $C$ & $A$  & $L(\rm{[OIII]})$\\
 (1)    & (2)      & (3) & (4) & (5) & (6)   & (7)                   & (8)     & (9) & (10) & (11) \\
\hline
1-211311 & 1-25688 & 261.284851 & 58.764687 & 0.0292 & -18.79 & 10.32 & S & 0.29 & 0.06 & 0.10$\pm$0.02\\
         & 1-94422 & 250.453201 & 41.818737 & 0.0316 & -19.15 & 10.55 & S & 0.37 & 0.05 & 0.24$\pm$0.03\smallskip\\
1-373161 & 1-259650 & 196.611053 & 45.289001 & 0.0509 & -21.07 & 11.68 & E & 0.44 & 0.06 & 0.67$\pm$0.20\\
         & 1-289865 & 322.048584 & 0.299885 & 0.0525 & -20.90 & 11.35 & -- & 0.49 & 0.02 & 0.11$\pm$0.09\\
1-210646 & 1-114306 & 323.742737 & 11.296529 & 0.0637 & -20.58 & 10.83 & S & 0.26 & 0.05 & 0.33$\pm$0.16\\
         & 1-487130 & 164.447296 & 21.233431 & 0.0587 & -20.47 & 10.86 & S & 0.26 & 0.11 & 0.27$\pm$0.10\smallskip\\
1-351790 & 1-23731 & 260.746704 & 60.559292 & 0.0205 & -18.20 & 10.19 & E & 0.40 & 0.02 & 0.02$\pm$0.01\\
         & 1-167334 & 151.894836 & 46.093983 & 0.0243 & -18.89 & 10.60 & E & 0.43 & 0.04 & 0.47$\pm$0.05\\
1-163831 & 1-247456 & 232.823196 & 45.416538 & 0.0705 & -20.05 & 10.83 & -- & 0.40 & 0.02 & 0.57$\pm$0.16\\
         & 1-210593 & 244.419754 & 41.899155 & 0.0605 & -19.76 & 10.90 & S & 0.36 & 0.06 & 0.43$\pm$0.14\smallskip\\
1-22301 & 1-251871 & 214.506760 & 41.827644 & 0.1027 & -21.17 & 11.68 & S & 0.26 & 0.05 & 0.24$\pm$0.18\\
         & 1-72914 & 127.580818 & 45.075867 & 0.0970 & -20.88 & 11.31 & S & 0.23 & 0.08 & 0.13$\pm$0.07\\
1-248420 & 1-211063 & 247.058411 & 40.313835 & 0.0331 & -19.87 & 10.79 & S & 0.33 & 0.09 & 0.20$\pm$0.04\\
         & 1-211074 & 247.462692 & 39.766510 & 0.0318 & -19.71 & 10.79 & S & 0.30 & 0.18 & 0.20$\pm$0.04\smallskip\\
1-23979 & 1-320681 & 213.813095 & 47.873344 & 0.0279 & -18.76 & 10.77 & E & 0.48 & 0.01 & 0.09$\pm$0.07\\
         & 1-519738 & 206.514709 & 22.118843 & 0.0277 & -19.49 & 10.73 & E & 0.45 & 0.03 & 0.11$\pm$0.04\\
1-542318 & 1-285052 & 199.061493 & 47.599365 & 0.0573 & -19.77 & 10.85 & S & 0.32 & 0.04 & 0.11$\pm$0.03\\
         & 1-377125 & 112.221359 & 41.307812 & 0.0585 & -19.67 & 10.84 & S & 0.41 & 0.02 & 0.57$\pm$0.14\smallskip\\
1-95092 & 1-210962 & 246.358719 & 39.870697 & 0.0290 & -20.49 & 11.09 & S & 0.47 & 0.07 & 0.35$\pm$0.06\\
         & 1-251279 & 209.251984 & 43.362034 & 0.0329 & -20.11 & 10.97 & E & 0.47 & 0.04 & 0.37$\pm$0.06\\
1-279676 & 1-44789 & 120.890366 & 47.892406 & 0.0586 & -19.33 & 10.92 & -- & 0.31 & 0.13 & 0.32$\pm$0.09\\
         & 1-378401 & 117.904335 & 48.000526 & 0.0612 & -19.65 & 11.02 & E & 0.41 & 0.02 & 0.57$\pm$0.14\smallskip\\
1-201561 & 1-24246 & 264.840790 & 56.567070 & 0.0818 & -19.91 & 10.57 & S & 0.75 & 0.36 & 0.11$\pm$0.06\\
         & 1-285052 & 199.061493 & 47.599365 & 0.0573 & -19.77 & 10.85 & S & 0.32 & 0.04 & 0.11$\pm$0.03\\
1-198182 & 1-256185 & 165.568695 & 44.271709 & 0.0370 & -20.00 & 11.03 & E & 0.50 & 0.06 & 0.25$\pm$0.04\\
         & 1-48053 & 132.595016 & 55.378742 & 0.0308 & -20.24 & 11.49 & E & 0.50 & 0.01 & --\smallskip\\
1-96075 & 1-166947 & 147.335007 & 43.442989 & 0.0720 & -20.79 & 10.81 & S & 0.29 & 0.02 & 0.13$\pm$0.08\\
         & 1-52259 & 59.411037 & -6.274680 & 0.0678 & -20.69 & 11.12 & S & 0.23 & 0.07 & 0.30$\pm$0.09\\
1-519742 & 1-37079 & 42.092335 & 0.986465 & 0.0274 & -17.25 & 9.55 & E & 0.27 & 0.02 & 0.02$\pm$0.01\\
         & 1-276679 & 161.272629 & 44.054291 & 0.0253 & -18.27 & 10.10 & S & 0.24 & 0.03 & 0.05$\pm$0.01\smallskip\\
1-491229 & 1-94554 & 248.914688 & 42.461296 & 0.0318 & -18.96 & 10.57 & S & 0.55 & 0.07 & 0.22$\pm$0.04\\
         & 1-604048 & 50.536137 & -0.836265 & 0.0365 & -20.37 & 10.91 & S & 0.42 & 0.09 & 0.39$\pm$0.08\\
1-604761 & 1-210173 & 241.341766 & 42.488312 & 0.0778 & -20.71 & 11.10 & S & 0.33 & 0.07 & 0.52$\pm$0.13\\
         & 1-71525 & 118.344856 & 36.274380 & 0.0457 & -20.17 & 10.97 & S & 0.27 & 0.10 & 0.19$\pm$0.06\smallskip\\
1-25725 & 1-211079 & 247.438034 & 39.810539 & 0.0304 & -18.97 & 10.54 & E & 0.54 & 0.01 & 0.03$\pm$0.04\\
         & 1-322074 & 228.700729 & 43.665970 & 0.0274 & -18.15 & 10.10 & E & 0.45 & 0.02 & --\\
1-94604 & 1-295095 & 248.348663 & 24.776577 & 0.0410 & -18.40 & 10.14 & E & 0.35 & 0.05 & 0.15$\pm$0.03\\
         & 1-134239 & 241.416443 & 46.846561 & 0.0571 & -19.83 & 10.70 & S & 0.36 & 0.04 & 0.23$\pm$0.06\smallskip\\
1-37036 & 1-210785 & 246.765076 & 39.527386 & 0.0338 & -20.22 & 10.97 & E & 0.47 & 0.01 & --\\
         & 1-25680 & 261.968872 & 60.097275 & 0.0278 & -19.41 & 10.84 & E & 0.52 & 0.02 & 0.34$\pm$0.04\\
1-167688 & 1-235587 & 214.854660 & 45.864250 & 0.0267 & -18.88 & 10.48 & E & 0.44 & 0.01 & 0.08$\pm$0.02\\
         & 1-37062 & 41.846367 & 0.058757 & 0.0248 & -18.30 & 10.40 & E & 0.49 & 0.03 & 0.27$\pm$0.03\smallskip\\
1-279666 & 1-392976 & 156.428894 & 37.497524 & 0.0432 & -17.91 & 10.09 & E & 0.37 & 0.02 & 0.10$\pm$0.03\\
         & 1-47499 & 132.037582 & 54.309921 & 0.0461 & -18.53 & 10.51 & E & 0.27 & 0.04 & 0.15$\pm$0.06\\
1-339163 & 1-136125 & 254.044144 & 34.836521 & 0.0316 & -19.33 & 10.50 & S & 0.25 & 0.09 & 0.08$\pm$0.02\\
         & 1-626830 & 204.683838 & 26.328539 & 0.0282 & -19.23 & 10.67 & S & 0.28 & 0.07 & 0.15$\pm$0.04\smallskip\\
1-258774 & 1-379660 & 119.973717 & 55.374817 & 0.0357 & -19.44 & 10.74 & E & 0.47 & 0.03 & 0.37$\pm$0.07\\
         & 1-48208 & 134.008118 & 57.390965 & 0.0406 & -19.57 & 10.85 & S & 0.50 & 0.01 & 0.12$\pm$0.04\\
1-198153 & 1-211063 & 247.058411 & 40.313835 & 0.0331 & -19.87 & 10.79 & S & 0.33 & 0.09 & 0.20$\pm$0.04\\
         & 1-135810 & 250.123138 & 39.235115 & 0.0297 & -19.38 & 10.59 & S & 0.24 & 0.09 & 0.08$\pm$0.02\smallskip\\
1-91016 & 1-338828 & 115.641609 & 44.215858 & 0.0418 & -18.10 & 10.42 & S & 0.28 & 0.03 & 0.43$\pm$0.05\\
         & 1-386695 & 137.983505 & 27.899269 & 0.0474 & -19.33 & 10.48 & S & 0.27 & 0.09 & 0.81$\pm$0.09\\
1-279073 & 1-211100 & 247.830322 & 39.744129 & 0.0309 & -19.15 & 10.62 & E & 0.56 & 0.02 & --\\
         & 1-210784 & 247.097122 & 39.570305 & 0.0292 & -19.61 & 10.86 & E & 0.48 & 0.00 & 0.15$\pm$0.05\smallskip\\
1-135044 & 1-218280 & 124.003311 & 27.075895 & 0.0255 & -19.57 & 10.81 & S & 0.27 & 0.08 & 0.12$\pm$0.03\\
         & 1-211063 & 247.058411 & 40.313835 & 0.0331 & -19.87 & 10.79 & S & 0.33 & 0.09 & 0.20$\pm$0.04\\
1-148068 & 1-166947 & 147.335007 & 43.442989 & 0.0720 & -20.79 & 10.81 & S & 0.29 & 0.02 & 0.13$\pm$0.08\\
         & 1-55572 & 133.121307 & 56.112690 & 0.0454 & -20.11 & 11.03 & S & 0.40 & 0.06 & 0.12$\pm$0.04\\
\hline
\end{tabular}
\end{table*}

\begin{table*}
\contcaption{}
\begin{tabular}{ccccccccccccc}
\hline
AGN mangaID & mangaID & RA  & DEC & $z$ & $M_r$ & log $M^\star/M_\odot$ & GZ1$_c$ & $C$ & $A$  & $L(\rm{[OIII]})$\\
 (1)    & (2)      & (3) & (4) & (5) & (6)   & (7)                   & (8)     & (9) & (10) & (11) \\
\hline
1-277552 & 1-264513 & 236.941513 & 28.641697 & 0.0333 & -20.92 & 11.28 & S & 0.25 & 0.18 & 0.33$\pm$0.05\\
         & 1-136125 & 254.044144 & 34.836521 & 0.0316 & -19.33 & 10.50 & S & 0.25 & 0.09 & 0.08$\pm$0.02\\
1-217050 & 1-135372 & 250.116714 & 39.320118 & 0.0301 & -20.29 & 11.08 & E & 0.49 & 0.02 & 0.01$\pm$0.23\\
         & 1-274663 & 157.660522 & 44.012722 & 0.0280 & -19.88 & 11.00 & E & 0.50 & 0.01 & 0.08$\pm$0.02\smallskip\\
1-25554 & 1-135625 & 248.507462 & 41.347946 & 0.0284 & -19.06 & 10.56 & S & 0.43 & 0.05 & 0.56$\pm$0.04\\
         & 1-216958 & 136.200287 & 40.591721 & 0.0270 & -18.95 & 10.41 & S & 0.51 & 0.03 & 0.23$\pm$0.02\\
1-135285 & 1-633990 & 247.304123 & 41.150871 & 0.0296 & -19.06 & 10.46 & S & 0.34 & 0.03 & 0.25$\pm$0.03\\
         & 1-25688 & 261.284851 & 58.764687 & 0.0292 & -18.79 & 10.32 & S & 0.29 & 0.06 & 0.10$\pm$0.02\\
\hline
\end{tabular}
\end{table*}

\section{Stellar populations}

In order to characterize the stellar population of the inner $3\times 3$\,arcsec$^2$ of the galaxies, we performed stellar population synthesis using the {\sc
starlight} code, developed by \citet{cid05}. This code combines, in different
proportions, $N_{\star}$ single stellar populations (SSPs) in
order to reproduce a galaxy observed spectrum -- excluding emission lines --, $O_{\lambda}$,  with a
model spectrum, $M_{\lambda}$. Each combination is found by solving
the equation
\begin{equation}
 M_{\lambda} = M_{\lambda_{0}} \left[ \displaystyle
\sum_{j=1}^{N_{\star}} x_{j}  b_{j,\lambda} r_{\lambda} \right]
\otimes G(v_{\star},\sigma_{\star}),
\end{equation}
were $b_{j,\lambda}\,r_{\lambda}$ is the reddened spectrum of the
$j$-th SSP normalized at
$\lambda_0$; $r_{\lambda}=10^{-0.4(A_{\lambda}-A_{\lambda 0})}$ is the
reddening term; $M_{\lambda 0}$ is the
synthetic flux at the normalisation wavelength; $\vec{x}$ is the
population vector.  The symbol $\otimes$ denotes the convolution
operator and $G(v_{\star},\sigma_{\star})$ is the Gaussian
distribution used to model the line-of-sight stellar
motions; it is centred at velocity $v_{\star}$  with dispersion
$\sigma_{\star}$. The reddening law we have used is that presented by
\citet{cardelli98}. The adopted normalization wavelength was $\lambda_0$ =5700\AA. This wavelength, usually selected to avoid emission and absorption features,
is similar to those adopted on other spectral synthesis studies, allowing a comparison of our results with previous ones.

The best fit model is determined by minimizing (through a simulated
annealing plus Metropolis scheme) the equation
\begin{equation}
\chi^2 = \sum_{\lambda}[(O_{\lambda}-M_{\lambda})w_{\lambda}]^2,
\end{equation}
where emission lines and spurious features are masked by assigning
$w_{\lambda}$=0 to these regions.

The spectral basis comprise the ``standard" 45 elements set, which is a sample
of the Evolutionary Population Synthesis models presented by
\citet{bc03}. It covers 15 ages: 0.001, 0.0031, 0.005, 0.01, 0.025,
0.04, 0.10, 0.28, 0.64, 0.94, 1.43, 2.5, 5.0, 11.0, 13.0 Gyr, and
three different metallicities:
$0.004\,Z_{\odot},0.02\,Z_{\odot},0.05\,Z_{\odot}$.

Since we are fitting the spectra of AGN, the signature of the central engine cannot
be ignored. This component was represented by a featureless continuum
(FC)
of power-law form that follows the expression $F_{\lambda}\propto
\lambda^{-0.5}$ \citep[see][for example]{koski78,riffel+09}.

\subsection{Results for the stellar population}

Running the STARLIGHT code allows the measurement of the relative contribution of each SSP vector, characterized by age and metallicity, 
to the total stellar light at 5700\AA{} ($x_j$) and its corresponding contribution to the total stellar mass ($\mu_j$),
that are the main quantities required to characterize the star formation history of the center of these galaxies.
The individual contribution of each SSP allows the derivation of the contribution of the main age and metallicity components to the total flux at 5700\AA,
although there are other parameters that globally represent the age and metallicity of the stars:
the mean light-weighted and mass-weighted ages ($\langle t_L \rangle$ and $\langle t_M\rangle$) and the light-weighted and mass-weighted metallicities \citep[$\langle Z_L \rangle$ and $\langle Z_M \rangle$,][]{cid05}.
The mean age, weighted by light and mass are, respectively, given by
\begin{eqnarray}
 \displaystyle \langle t_L\rangle = \displaystyle \sum_{j=1}^{N_{\star}} x_j \log t_j\ &  &  \langle t_M \rangle = \sum_{j=1}^{N_{\star}} \mu_j \log t_j\text{,}
\end{eqnarray}
\noindent and the mean metallicities, weighted by light and mass, are expressed by
\begin{eqnarray}
 \displaystyle\langle Z_L\rangle = \displaystyle \sum_{j=1}^{N_{\star}} x_j Z_j\ &  & \langle Z_M \rangle = \sum_{j=1}^{N_{\star}} \mu_j Z_j\text{,}
\end{eqnarray}
\noindent  respectively.

A strong contribution of one single 
SP age tends to be spread over different SSP components
of similar ages \citep{cid01}; thus, one must be must be careful interpreting 
a single SSP component. In order to obtain 
a more  robust and reliable description of the SFH, the  SSPs are usually grouped
in age bins \citep[e.g.][]{Dametto}, which we have chosen as follows: 
\emph{young-young}: $x_{yy}$ ($t\leq 10\,\times 10^{6}$\,yr),
\emph{young-old}: $x_{yo}$ ($25 \leq t \leq 40\,\times 10^{6}$ yr), 
\emph{intermediate-young}: $x_{iy}$ ($0.100 \leq t \leq 0.287\,\times 10^9 $ yr),
\emph{intermediate-intermediate}: $x_{ii}$ ($ 0.640 \leq t \leq 0.905\,\times 10^9$ yr),
\emph{intermediate-old}: $x_{io}$ ($1.0 \leq t \leq 2.5\,\times 10^9$ yr) and 
\emph{old}: $x_{o}$ ($4.0\leq t \leq13.1\,\times 10^9$ yr).

\begin{figure*}
 \centering
 \includegraphics[width=\textwidth]{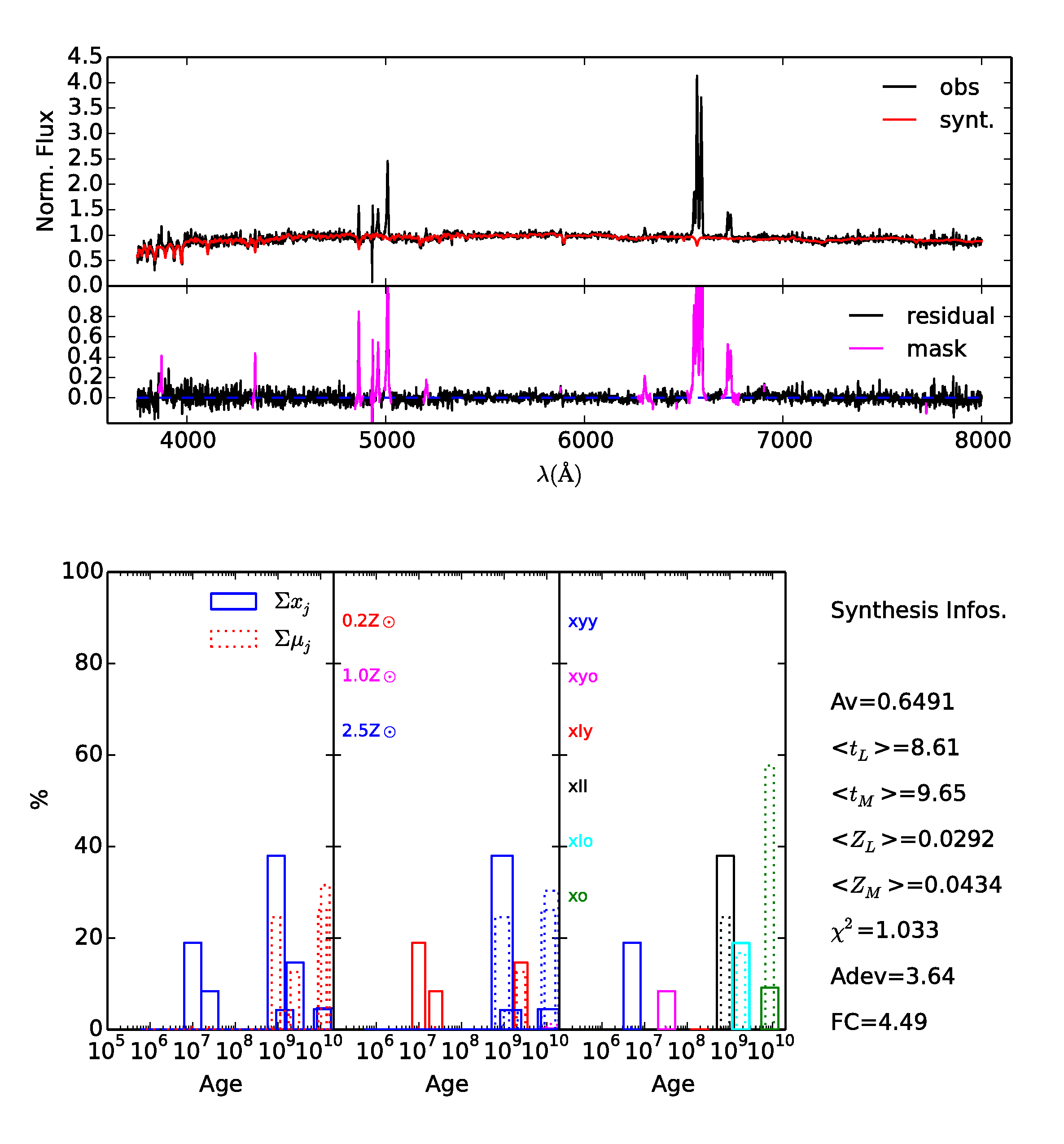}
\caption{Results from the stellar population synthesis for the strong AGN mangaID 1-269632. \emph{Upper two panels}: top -- observed (black) and synthetic (red) spectrum; bottom -- residuals (observed spectrum - synthetic), with the masked
regions highlighted in magenta. \emph{Bottom three panels}:
histograms where solid lines represent the light weighted contribution of the SSPs and the dashed lines represent the
mass-weighted contribution. The left histogram shows the individual contribution of each SSP, sorting the SSP vectors by age and summing the metallicities;
the middle histogram displays the same as the left histogram but separating the contribution of each SSP in metallicity, with different colors indicating different metallicities;
the right histogram presents the mass contributions of the SSPs divided in six age bins (see text), and summing the metallicities. To the right and bottom are presented the derived average ages and metallicities (both mass- and light-weighted), the extinction $A_V$, the contribution of the featureless continuum in percent units, the chi-squared value of the fit, and the average of the absolute percent difference between observed and synthesized spectra, Adev, that parameterizes the quality of the fit.}
\label{synthesis_exe}
 \end{figure*}

Figure \ref{synthesis_exe} presents an example of the results of the stellar population synthesis for the strong AGN mangaID 1-269632.
The upper panels show the observed and synthetic spectrum as well as the corresponding residuals; the bottom panels display the resulting light (percent contribution of each age bin above to the light at 5700\AA) and mass contributions 
of single and binned SSPs vectors. Also shown are the mean ages and metallicities.
The inner $\sim$3.7\,kpc of this galaxy  is dominated by the contribution of intermediate age stars ($x_{ii}\approx 38\%$ and $x_{io}= 19\%$) with
old stars contributing $x_{o} \approx 9\%$ and young stars $x_{yy}\approx 19\%$, which implies that 
at least one episode of star formation has occurred in the last 10 Myr. Tables with the synthesis results for all galaxies, both for the AGN and control samples, showing the contribution of the six age bins listed above (both in percent flux at 5700\AA{} and in mass) are presented in the Appendix.

Figure~\ref{stellar_light} summarizes the results from the synthesis in histograms of the percent contribution of each age bin to the continuum at 5700\AA{}. The results for the AGN sample are shown in blue for the ``weak AGN'' ($L(\rm{[OIII]})<3.8\times 10^{40}$\,erg\,s$^{-1}$), in green for the strong AGN and red for the control sample. In these histograms, we have chosen to present the results in bins of 10\% for the 3 oldest age ranges and in bins of 3\% for the three youngest ages ranges. We have adopted these smaller bins for the youngest ages to better sample the age distributions, as these ages never contribute more than 30\% to the total flux at 5700\AA{}, allowing to restrict the $x$-axis to a maximum contribution of 30\%.

\section{Discussion}

\subsection{Comparison between the AGN hosts and control sample}

\begin{figure*}
\centering
\subfloat[]{\label{f1}\includegraphics[width=\columnwidth]{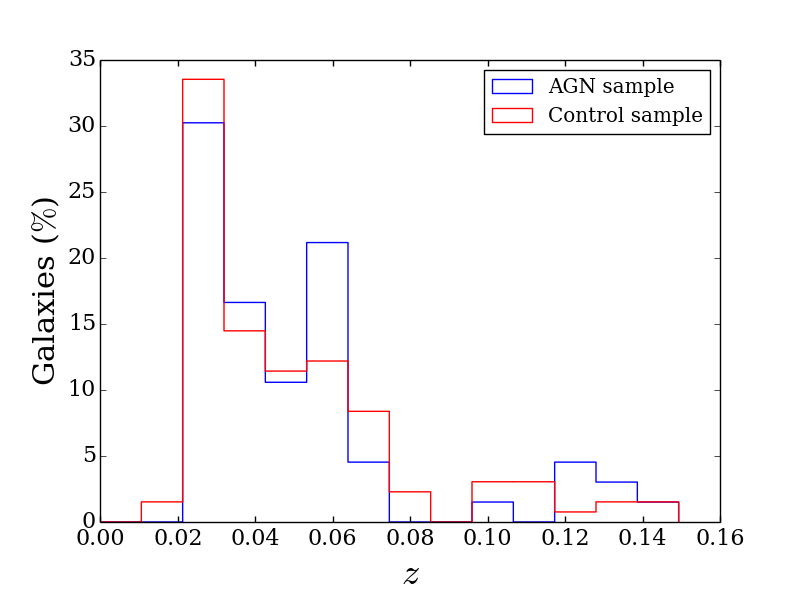}}
\subfloat[]{\label{f2}\includegraphics[width=\columnwidth]{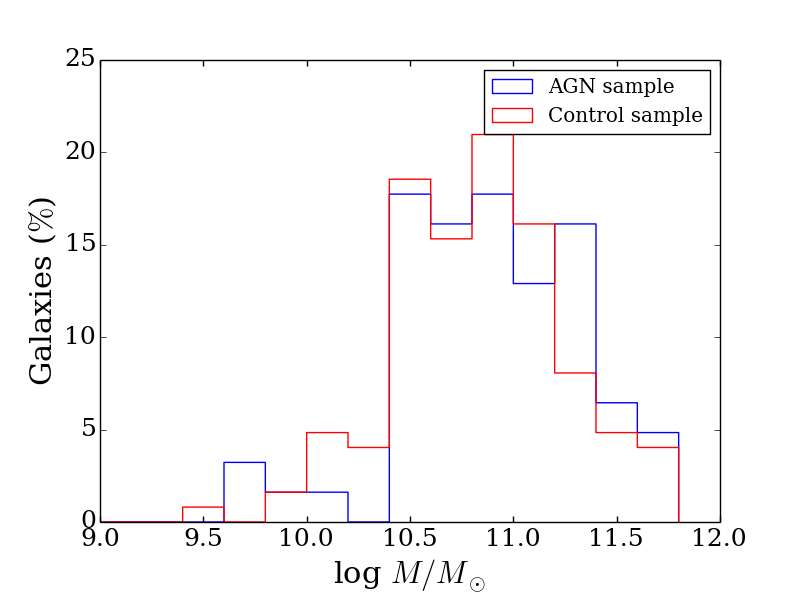}}\\
\subfloat[]{\label{f3}\includegraphics[width=\columnwidth]{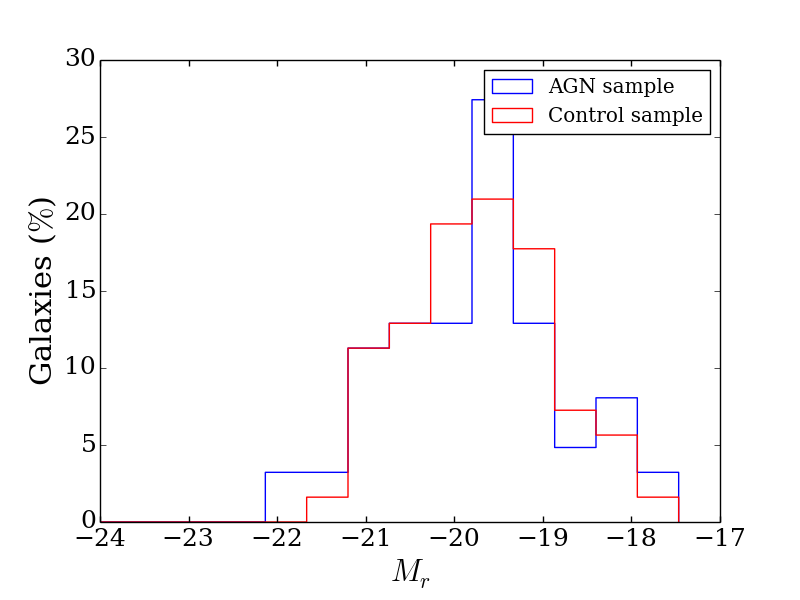}}
\subfloat[]{\label{f4}\includegraphics[width=\columnwidth]{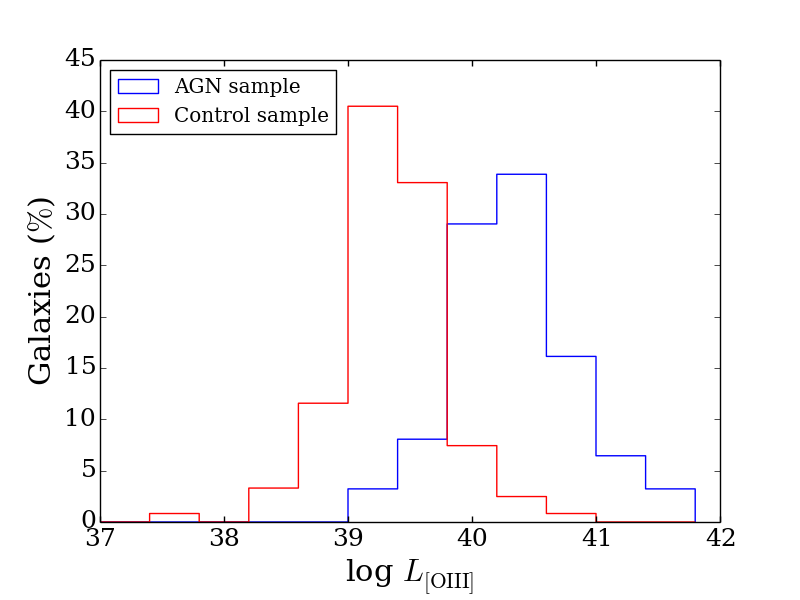}}
\caption{Distributions of redshift (a), stellar mass (b), absolute $r$-band magnitude (c) and [OIII] luminosity (d) of the AGN
host sample (blue) and the control sample (red). The AGN and the control sample present
markedly different distributions of $L(\rm{[OIII]})$, but have the other parameters similar between the two samples.}
\label{figdistr}
\end{figure*}

After selecting the control sample objects according to the criteria described in Sect.~\ref{sec:cs}, we now check if the resulting
distribution of stellar masses and redshifts are indeed compatible with that of the AGN hosts.
In Figure~\ref{figdistr} we show the distributions of stellar mass, $r$-band absolute magnitude and redshifts of the AGN sample and the control sample.
The distributions of stellar mass and $r$-band absolute magnitude of the two samples are very similar. The probability that these distributions for the AGN and non-AGN galaxies are not drawn from the same distribution is less than 5 per cent as indicated by an Anderson-Darling test (hereafter A-D test). Regarding the redshift distribution, the two samples are also similar; the statistical significance of the A-D test is lower (16 per cent), but this is not critical to our analysis.
The only significant difference observed in Figure~\ref{figdistr} is for the distribution of the [OIII] luminosity $L(\rm{[OIII]})$, as expected: while the centroid of the distribution for the AGN sample is $\log\,L(\rm{[OIII]})\approx 40.5$, for the control sample the centroid is $\log\,L(\rm{[OIII]})\approx 39.3$.

\begin{figure}
\includegraphics[width=\columnwidth]{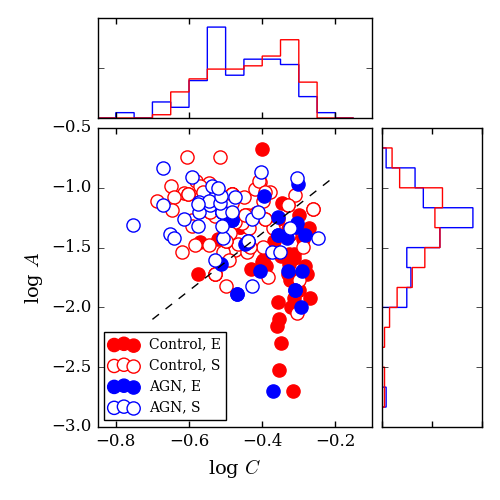}
\caption{Concentration $C$ and asymmetry $A$ indices values for AGN and the control sample galaxies. The Galaxy Zoo I morphological classification as ellipticals or spirals is represented by filled and open circles, respectively. Blue (red) circles represents AGN (control sample).
The dashed line roughly separates the spiral and elliptical-dominated regions in the diagram.
The distributions of $A$ and $C$ are similar, but the AGN present a small tendency to be
more concentrated near the dashed line.}
\label{ca_comb}
\end{figure}

We have also made a quantitative assessment of how well the control sample morphological distribution matches that of the AGN hosts.
The GZ1 classification reveals that the AGN sample contains 34 (55 per cent) spiral and 18 (29 per cent) elliptical galaxies. The remaining 10 objects (16 per cent) comprise 6 E/S galaxies, 1 merger and 3 unclassified objects. The morphological distribution of the control sample is the same within the
uncertainties: 60 per cent spiral and 34 per cent elliptical galaxies. Figure~\ref{ca_comb} displays the concentration and asymmetry distributions of the AGN and the control sample galaxies. The dashed line is the optimal separation between
the GZ1 major classes (ellipticals and spirals) in the diagram. The code PyCA does not calculate uncertainties in $C$ and $A$, but the good separation between elliptical and spiral
galaxies in the $C-A$ plane, with elliptical galaxies ocuppying the locus of low asymmetry and high concentration, suggests that the estimates of these parameters are robust.
The distributions of AGN hosts and non-active galaxies in the $C-A$ plane are similar, although a small systematic effect is discernible in the sense of a wider distribution of the control sample galaxies in the $C-A$ diagram relative to the AGN. This behaviour is produced by the fact that AGN galaxies appear to be less common in the extremes of low concentration / low asymmetry indices. This result suggests that the AGN sample is morphologically less diverse than the control sample, preferring to populate the intermediate region between ellipticals and spirals. This trend is, however, secondary, as the A-D test results in a $p$-value larger than 0.2 for the null hypothesis that the $C$ and $A$
distributions for AGN and control sample are drawn from the same distribution.

The combined SDSS-III \textit{ugriz} images of the four strongest AGN from our sample and their ``control partners'' are shown in Figure~\ref{agnimages}. Those for the rest of the sample are shown in the Appendix. Both figures reveal an apparent very good match in the morphology and axial ratios between the AGN and control sample.

\subsection{Stellar population properties}

\begin{figure*}
\centering
\includegraphics[width=0.8\textwidth]{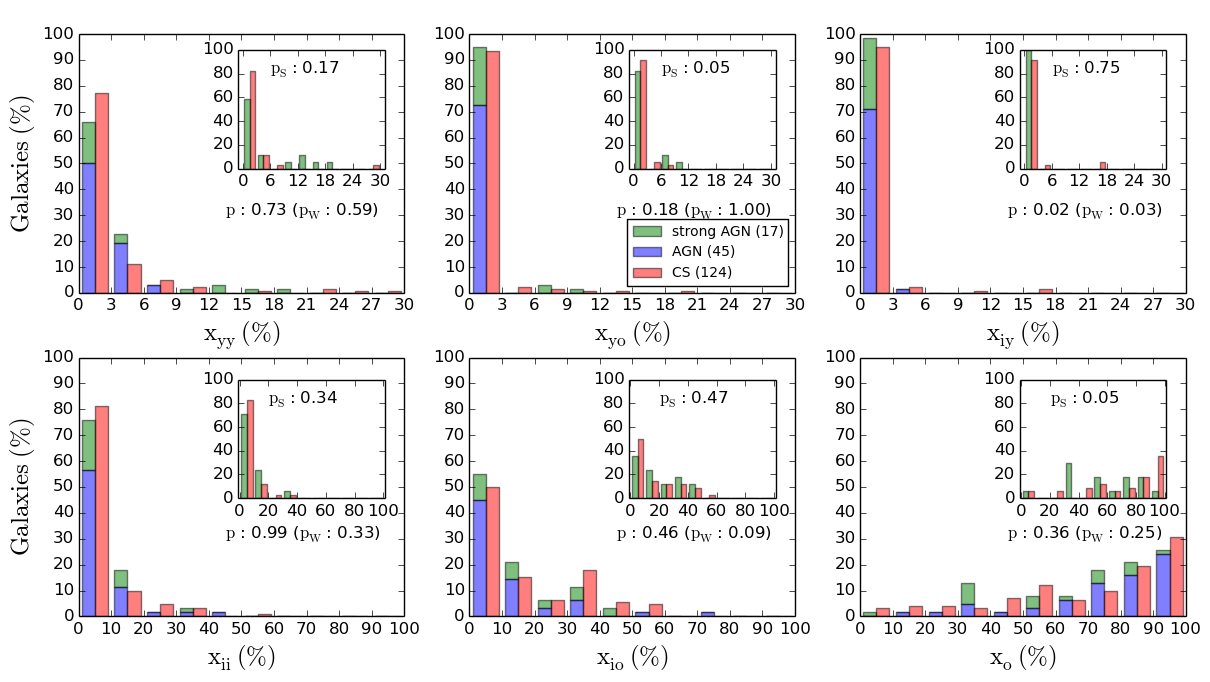}
\caption{Histograms of the distribution of galaxies according to the fractional contribution of each age bin to the total light at restframe 5700\,\AA{}.
Weak AGN are represented in blue, strong AGN in green and control sample in red.
For the age bins $x_{ii}$, $x_{io}$ and $x_{o}$ we show, in an insert,
histograms comparing only the strong AGN (green) with its respective control sample (red).
For each histogram the $p$-value of the A-D test is given. We also present, below the insert in each plot, the p-value of the comparison between the weak AGN and the control sample. $p_w$ refers to the weak AGN, $p_s$ refer to the strong AGN and $p$ to the combined sample of strong and weak AGN.}
\label{stellar_light}
\end{figure*}

Figure~\ref{stellar_light} shows that there is no clear difference between the stellar population of all AGN -- considering both the weak and strong AGN together -- and the control sample. But, when comparing only strong AGN with their control sample, the distributions differ at the 2-$\sigma$ level for $x_{yo}$ and $x_{o}$. Though the differences between the distributions are not 
obvious in these plots due to the histogram binning, they can be seen with smaller bin widths and are confirmed by the results of the A-D tests. The numbers in the top-right corners of each panel in Fig.~\ref{stellar_light} give the probability that the two distributions are derived  from the same sample. The small values for $x_{yo}$ and $x_o$ indicate that the contribution of these populations to the light at 5700\AA{} differ between the strong AGN and the control sample, in the sense that the strong AGN show a smaller contribution from the old stellar population and a larger contribution from the young-old component than the control sample.

The above results suggest that, when comparing weak and strong AGN,
the contribution of old stellar populations decreases, while that of the younger stellar populations
increases in the latter. This trend agrees with the previous results by \citet{kauffmann03}, who compared 
the stellar population properties of AGN-host galaxies and normal galaxies for a sample of $\sim$22600 galaxies with SDSS-I spectra
\citep{york00}.
These authors studied the mean stellar age and stellar formation history
of these AGN by measuring the indices $D_n(4000)\times W(H\delta)$ \citep{Balogh99}. 
They found that weak AGN have mostly old stellar populations that are similar to those of early-type galaxies (non-AGN), while the strong AGN have much younger stellar ages and typically strong H$\delta$ absorption-line equivalent width, indicating
that they have experienced a burst of star formation in the past 1-2 Gyr.
Our sample of strong AGN is comprised so far by 17 hosts, but our results are robust, as supported by the A-D tests, and will be increased as more of these objects are observed with MaNGA.

\begin{figure}
 \centering
 \includegraphics[width=\columnwidth]{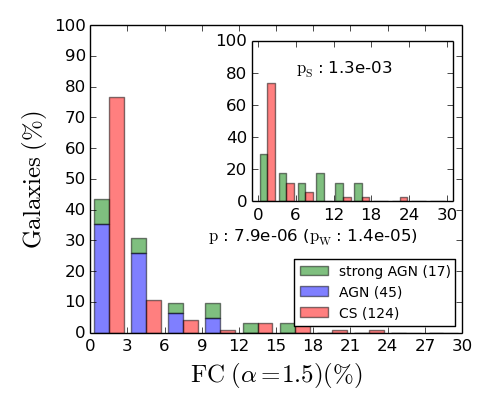}
\caption{Distribution of galaxies according to the fractional contribution of the featureless continuum to the total light at restframe 5700\,\AA{}.
Weak AGN are represented in blue, strong AGN in green and control sample in red.
In the insert, we show the histogram comparing only the strong AGN (green) with its respective control sample (red).
The $p$-values of the A-D test of the comparison between the AGN sample and the control sample are also shown. $p_w$ refers to the weak AGN, $p_s$ refer to the strong AGN and $p$ to the combined sample of strong and weak AGN.}
\label{fcfig}
 \end{figure}

The results of the synthesis for the FC continuum, presented in  Fig.~\ref{fcfig}, show a clear difference between the AGN and control sample: while the control sample tend to show very small contribution of FC in most cases, for the AGN, and in particular for the strong AGN, the FC contribution tends to be larger, as expected. In any case, this contribution is usually smaller than 10\%, and in only a few cases it reaches $\sim$\,20\%. But we also point out that a common problem of stellar population syntheses of the spectra of active galaxies  is that a reddened young starburst ($t\leq$5M\,yr) is very difficult to distinguish from that of an AGN continuum as discussed in previous studies \citep{cid95,thaisa00,cid04,riffel+09}. This is particularly true when the FC contribution is smaller than $\approx$\,20\%, as is the case here. We thus conclude that, although some degeneracy may be occurring, the fact that we are finding more FC contribution for the AGN than for the control sample supports that we are being able to separate this contribution in many cases.
In order to estimate the impact of this degeneracy on the derived population fractions we have performed a series of simulations.
We have combined a set of SSPs from our base with contributions similar to those we have found of our sample, and added a moderate FC contribution of 10\%.
We have then perturbed the individual flux values using the same error
distribution as the sample spectra, and run {\sc Starlight} with the same configuration as before, both allowing or
not allowing for an FC in the fit. We have found that, when the FC is present but the synthesis does not allow for an FC in the fit, the contributions
of young and intermediate stellar populations are overestimated by $\sim 5\%$ on average. However, when the FC is allowed,
the contribution of young stellar populations is only slightly overestimated ($\sim 2\%$). This agrees with the recent results of \citet{cardoso17} that have shown that, for a broad range in star formation histories, the effect of a FC in the derived mean stellar ages is typically lower than 0.1\,dex even for FC contributions as large as 40\%. Therefore, in the cases that there may be some degeneracy, the effect should be small, with no significant influence on the results of the synthesis discussed above.

\begin{figure*}
 \centering
\begin{tabular}[b]{c}
\includegraphics[width=0.7\textwidth]{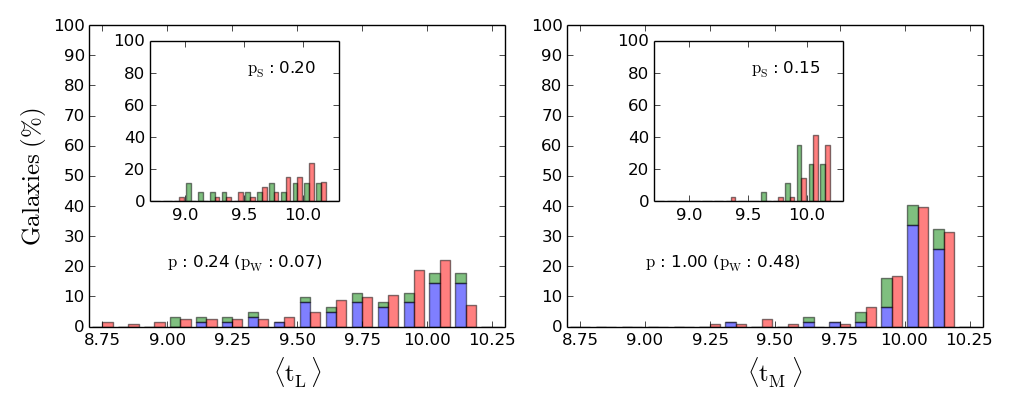}\\
\includegraphics[width=0.7\textwidth]{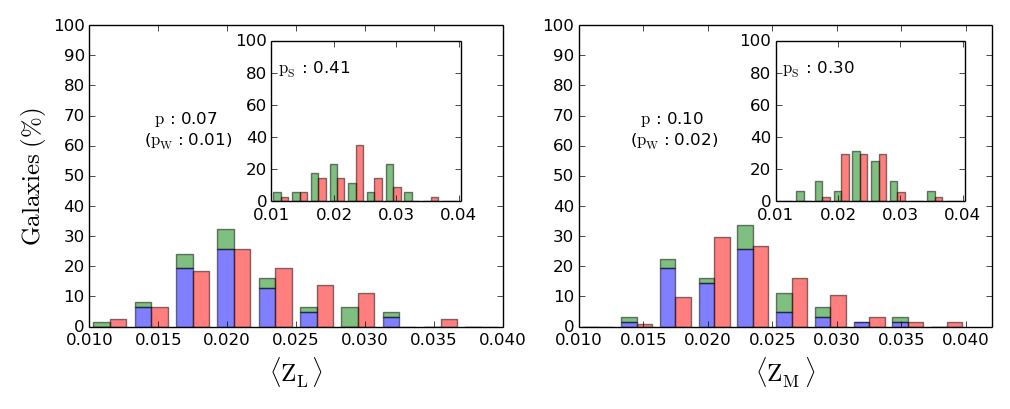}
\end{tabular}
\caption{Histograms showing the distributions of the mean ages (top panel) and metallicities (bottom panel), for the AGN (blue for weak and green for strong) and control sample (red). In the insert of each pannel, we plot the histograms comparing the strong AGN (green) with its respective control sample (red). For each histogram the $p$-value of the A-D test is given. $p_w$ refers to the weak AGN, $p_s$ refer to the strong AGN and $p$ to the combined sample of strong and weak AGN.}
\label{mean5700}
\end{figure*}

The distribution of the global parameters, mean ages $\langle t_L \rangle$ and $\langle t_M \rangle$ and mean metallicities  
$\langle Z_L \rangle$ and $\langle Z_M \rangle$ are presented in Figure~\ref{mean5700}.
This figure shows that the distribution of mean ages (weighted in light) differ at the 2-$\sigma$ level between the weak AGN and the control sample, while no such difference is observed for the strong AGN. On the other hand, if one calculates the median age for all weak and all strong AGN, one finds no difference relative to the controls for the weak AGN ($\langle t_L \rangle=9.90\pm0.25$) but a difference of 0.2\,dex (younger)  for the strong AGN ($\langle t_L \rangle=9.70\pm0.25$, with the controls presenting the same value as observed for the weak AGN). 

Regarding the metallicity, Fig.~\ref{mean5700} reveals a small difference between the weak AGN and their control sample, with a probability of only about 1\% that the distributions are similar in the case of normalization in light ($Z_L$), while no such difference is observed for the strong AGN. Although the difference in metallicity between the AGN hosts and control galaxies may be due to a different origin for the stars, there is also the possibility that we are observing the effect of degeneracy between age and metallicity in optical spectra \citep{Worthey}. This degeneracy arises because the lower metallicity leads to bluer stellar atmospheres, what is also the signature of young stars. As this effect is small, we defer its analysis to a future study, when we have a larger sample of AGN hosts.

\subsubsection{The effect of the AGN luminosity}

Although we have explored the effect of the AGN luminosity above by separating the analysis of the stellar population in only two populations  -- strong and weak AGN -- we now investigate in more detail this effect by looking for possible correlations of the stellar population properties with L[OIII]. We also take advantage of our careful selection of the control sample to explore the differences between the properties of each AGN host and its two controls, investigating how these differences may be associated to the AGN luminosity.

\begin{figure*}
 \centering
\includegraphics[width=0.9\textwidth]{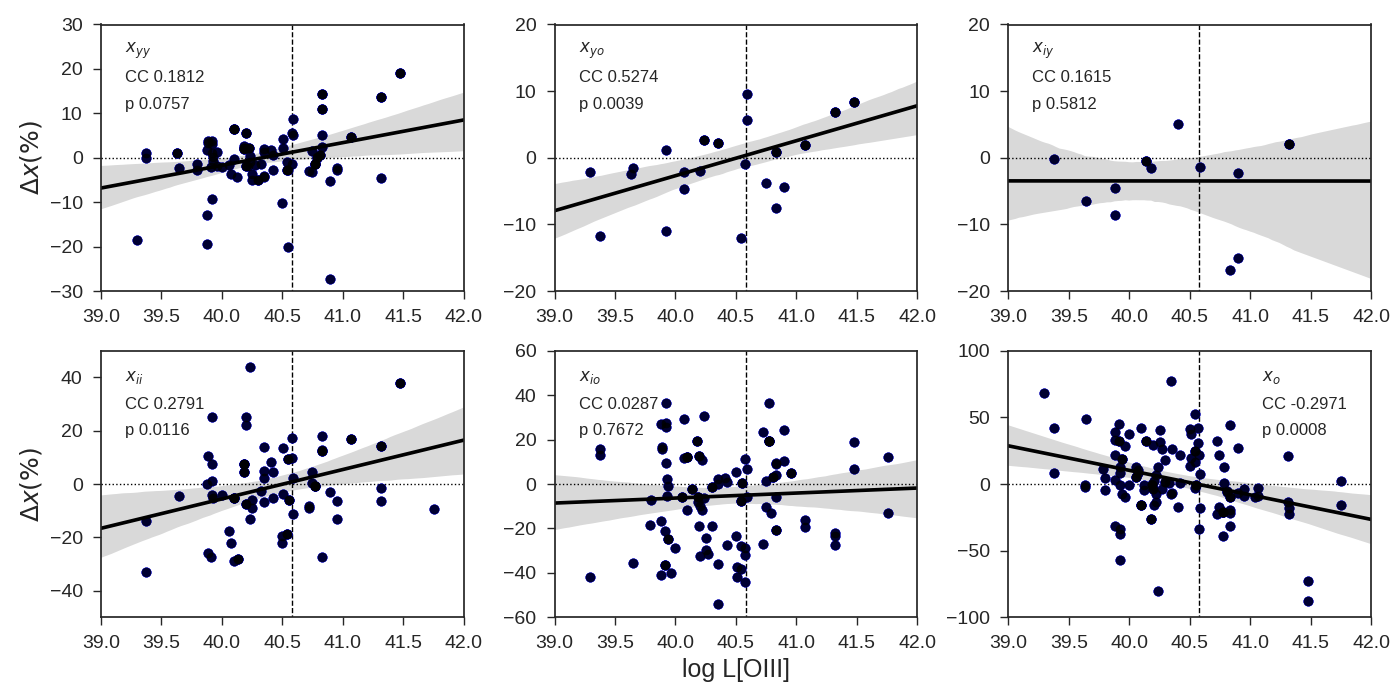}
\caption{Differences $\Delta x$ in the fractional contribution of each age bin to the total light at restframe 5700\,\AA{} between
AGN and control galaxies, as a function of the AGN [OIII] luminosity. Each circle correspond to a pair AGN / control object. We have excluded
pairs for which the fractional contributions are both zero. The best-fit linear regression and its uncertainties are shown with a thick line and
a shaded area, respectively. The Spearman correlation coefficient (CC) and the p-value of the test are given in each pannel. The [OIII] luminosity which separates strong and weak AGN is shown as a vertical dashed line. The locus $\Delta x=0$ is shown by a dotted line.}
\label{agebins_oiii}
\end{figure*}

For this exercise we calculated the differential fractional contributions (AGN - control): the differences $\Delta x$ (in each age bin $x$) of the fractional contributions to the light at 5700\AA\ between each AGN and its two control partners, and verified how, and if, it depends on the AGN [OIII] luminosity. The results are shown in Figure~\ref{agebins_oiii}, where the limit between strong and weak AGN is shown as a vertical dashed line. We have performed a linear regression between $\Delta x$ and L[OIII] for each age bin and the best fit is shown as a continuous line in the different panels of the figure.

There is an obvious correlation between $\Delta x$ and L[OIII] for some age bins. This is the case for $x_{yo}$: a Spearman test gives a correlation coefficient of 0.53 with a p-value of 0.0039. For the oldest age bin, $x_{o}$, there is an anti-correlation between $\Delta x$ and L[OIII], with a Spearman correlation coefficient of -0.30 and a statistical significance higher than 3-$\sigma$. Another result from this figure is that an AGN host
can have a higher or lower contribution of a given age bin, depending on the AGN luminosity. In the bin $x_{o}$, for example, the contribution is higher in lower-luminosity AGN 
than in their ``control partners'', while  the opposite is true for luminous AGN, which present a smaller contribution of this age bin than their respective control galaxies. A similar
behaviour can be seen in the bin $x_{yo}$: the relative contribution of this age bin for low-luminosity AGN is smaller, but for high-luminosity AGN is higher.

\begin{figure*}
 \centering
\includegraphics[width=0.7\textwidth]{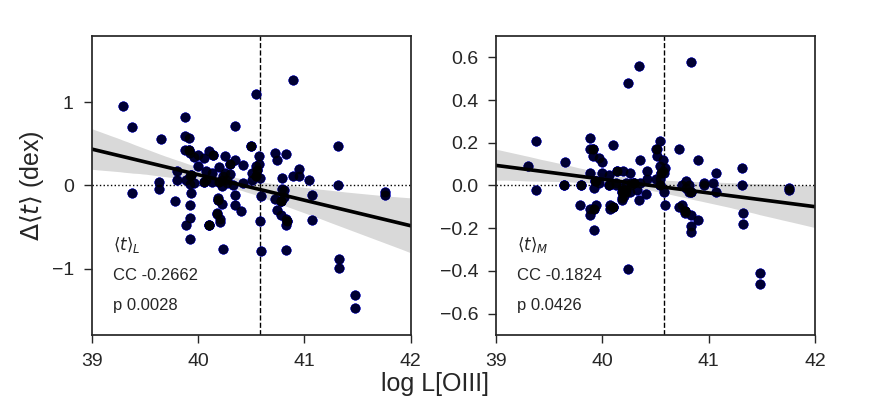}
\caption{Light-weighted (\textit{left}) and mass-weighted (\textit{right}) differences $\Delta \langle t\rangle$ in the mean stellar age between
AGN and control galaxies, as a function of the AGN [OIII] luminosity. The symbols are the same as in Figure~\ref{agebins_oiii}.}
\label{meanage_oiii}
\end{figure*}

We have also explored how the differences $\Delta \langle t \rangle$ in the (light-weighted and mass weighted) logarithmic mean stellar age between AGN and control sample (which we will refer to as ``differential age'') depends on the AGN luminosity. The result is shown in Figure~\ref{meanage_oiii}. It is evident that $\Delta \langle t\rangle$ and
L[OIII] show an inverse correlation, especially for the light-weighted mean ages. The resulting Spearman correlation coefficient is -0.27, with a statistical significance higher than 2-$\sigma$. The overall results agree with those obtained in Figure~\ref{agebins_oiii} and indicate that
luminous AGN are younger, on average, than inactive galaxies with similar properties. Low-luminosity AGN, on the contrary, present older
stellar populations than the control sample. Notice, also, that the separation between these two AGN categories is very close
to the threshold $L(\rm{[OIII]})=3.8\times 10^{40}$\,erg\,s$^{-1}$ (the vertical dashed line), which marks the separation between our weak and strong
AGN samples. These results agree with those from \citet{kauffmann03}, which showed that, at fixed stellar mass density, luminous AGN present younger stellar populations than inactive galaxies, while low-luminosity AGN are older, on average.

In oder to translate the above results in numbers, we show in Table\,\ref{global} the mean ages of the AGN and their control galaxies separated in bins of AGN luminosity. Although some bins still have a few galaxies (that we hope to increase as the survey progresses), and the dispersion in the mean age values is large, Table\,\ref{global} shows a steady decrease of the mean age of the host galaxy as the AGN luminosity increases, besides showing also the trend discussed regarding the difference in age between the AGN hosts and controls.

\begin{table*}
\caption{Average of global parameters $\langle t_L \rangle$ and $\langle t_M \rangle$, for the AGN and their control galaxies separated by luminosity.}
\begin{tabular}{lcccccc}
\hline
$\log_{10}\mathrm{L[OIII]}$ 	& N  	& \multirow{2}{*}{Sample} & $\langle t \rangle_{L}$ & $\Delta \langle t \rangle_{L}$  &  $\langle t \rangle_{M}$ & $\Delta \langle t \rangle_{M}$ \\
(erg\,s$^{-1}$) 		& (AGN) & {} & $\log_{10}$ (yr) & $\mathrm{dex}$(year) & $\log_{10}$ (yr) & $\mathrm{dex}$(year)\\
\hline
{} & {} & {} & {} & {} & {} & {} \\

\multirow{2}{*}{39.0 -- 39.75}	& \multirow{2}{*}{5}	& AGN		& 9.89$\pm$0.16 & \multirow{2}{*}{0.24}	& 10.07$\pm$0.03 &  \multirow{2}{*}{0.05}	\\
  {}				& {}			& Control	& 9.65$\pm$0.42 & {}			& 10.02$\pm0.10$ & {}				\\
{} & {} & {} & {} & {} & {} & {} \\

\multirow{2}{*}{39.75 -- 40.25}	& \multirow{2}{*}{25} 	& AGN		& 9.81$\pm$0.30 & \multirow{2}{*}{0.05}	& 10.00$\pm$0.17 &  \multirow{2}{*}{0.0}	\\
 {}				& {}			& Control	& 9.76$\pm$0.36 & {}			& 10.00$\pm$0.19 & {}				\\
{} & {} & {} & {} & {} & {} & {} \\

\multirow{2}{*}{40.25 -- 40.75}	& \multirow{2}{*}{18} 	& AGN		& 9.81$\pm$0.27 & \multirow{2}{*}{0.10}	& 10.05$\pm$0.06 &  \multirow{2}{*}{0.04}	\\
 {}				& {} 			& Control	& 9.71$\pm$0.32 & {}			& 10.01$\pm$0.12 & {}				\\
{} & {} & {} & {} & {} & {} & {} \\

\multirow{2}{*}{40.75 -- 41.25}	& \multirow{2}{*}{10} 	& AGN		& 9.66$\pm$0.35 & \multirow{2}{*}{-0.08}& 9.99$\pm$0.11	 &  \multirow{2}{*}{-0.01}	\\
 {}				& {}			& Control	& 9.74$\pm$0.43 & {}			& 10.00$\pm$0.18 & {}				\\
{} & {} & {} & {} & {} & {} & {} \\

 \multirow{2}{*}{41.25 -- 42.0}	& \multirow{2}{*}{4} & AGN	& 9.43$\pm$0.64 & \multirow{2}{*}{-0.55}	& 9.94$\pm$0.64 &  \multirow{2}{*}{-0.14}		\\
 {}			& {} & Control	& 9.98$\pm$0.15 & {}			& 10.08$\pm$0.03 & {}						\\
{} & {} & {} & {} & {} & {} & {} \\
\hline
\end{tabular}
\label{global}
\end{table*}

\subsection{Results from MaNGA data}

As pointed out in the Introduction, we are applying the same methodology described above to study the resolved stellar population in the MANGA data (Mallmann et al., in preparation, hereafter Paper II). The SDSS-III spectra correspond to a region of radius 1.5'' around the nucleus -- corresponding to a range of radii from $\approx$ 1 to 3 kpc at the distance of the galaxies. This is similar to the area covered by the central pixel of the MaNGA array. Thus we have checked, for a subsample of representative galaxies, whether our synthesis result for the inner pixel was consistent with that obtained from the SDSS-III spectra. We concluded that the results were the same within the uncertainties, as expected. 

\begin{landscape}
\begin{figure}
\flushleft
 \includegraphics[width=4.7in]{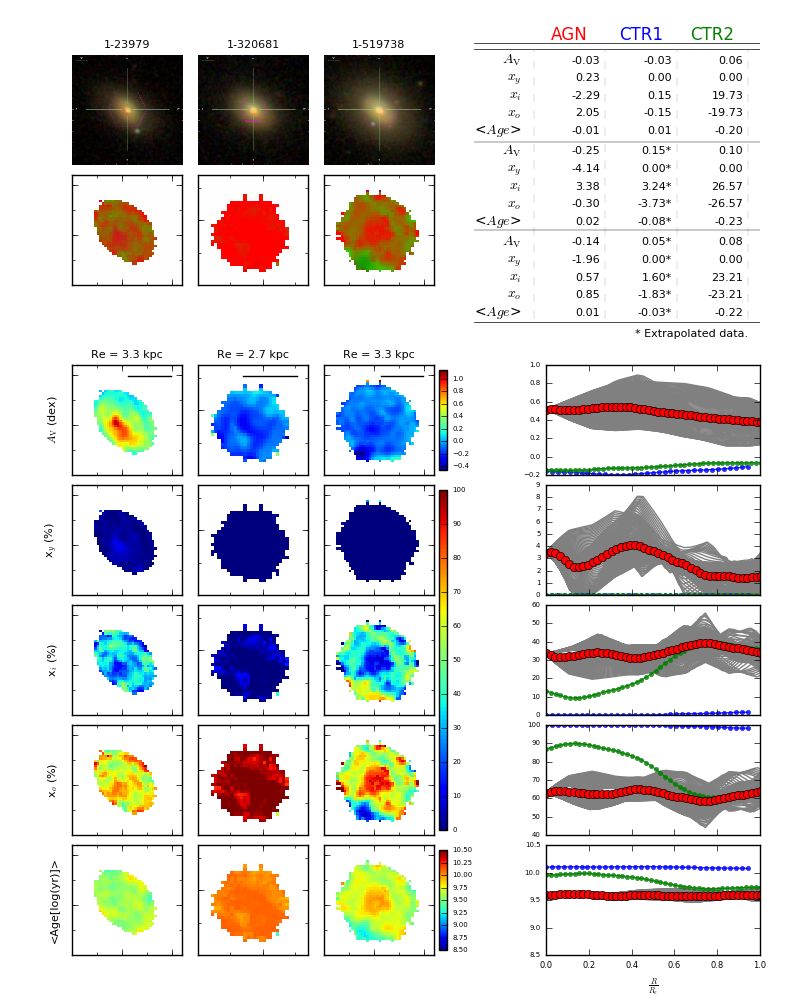}
 \includegraphics[width=4.7in]{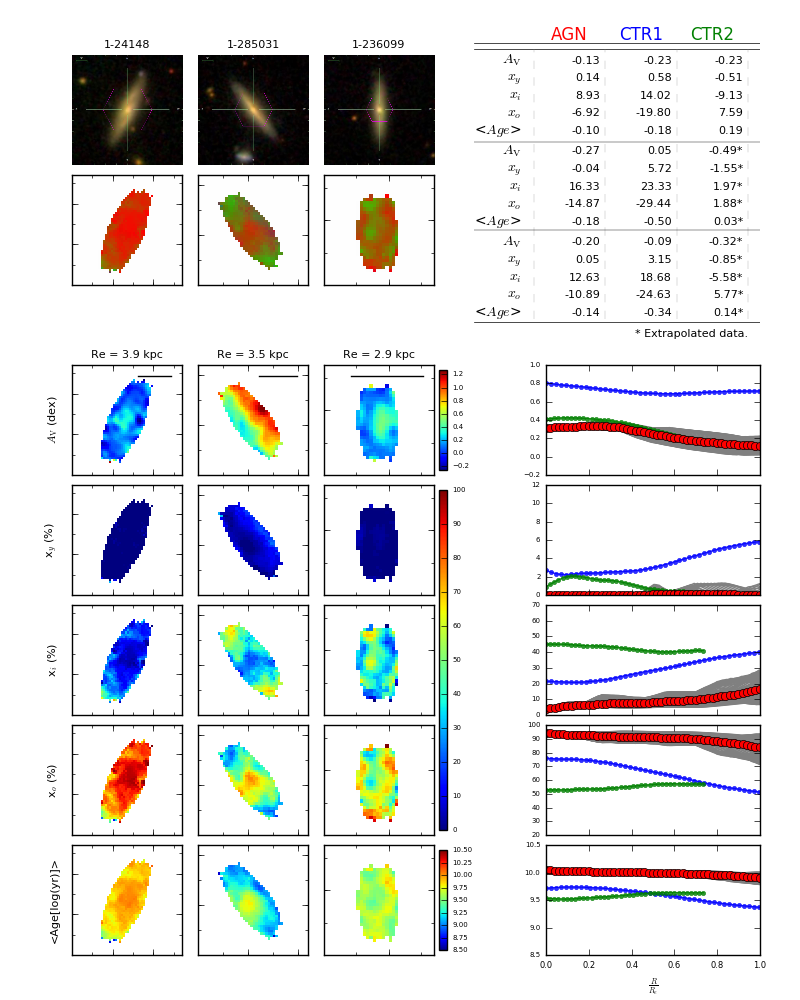}
 \caption{Comparison between the spatially resolved stellar population of an AGN (first column) and its two control galaxies 
 (second and third columns). {\bf Top left}: two rows of panels,  SDSS image of the galaxies (top) and RGB maps (bottom) 
 showing the relative percent contribution of the young ($< 40$ Myr, blue), intermediate age ($50Myr$ - $2Gyr$, green) 
 and old ($> 2Gyr$, red) stellar population. {\bf Top right}: table summarizing the average gradient values for the 
 different properties calculated at three effective radius ranges ($0.0-0.5 R_e$, $0.5-1.0 R_e$ and $0.0-1.0 R_e$) 
 using the mean profiles shown. {\bf Bottom}: five rows of panels, from top to bottom: visual extinction $A_{\rm v}$, 
 percent contribution of the young (x$_y$), intermediate age (x$_i$) and old (x$_o$), and average age ($<Age>$). 
 Fourth column: mean profiles of the properties shown in the left figures, as a function of effective radius. 
 Active galaxy in red, control galaxies in blue and green. Grey lines show the profiles of the active galaxy 
 for each sector of 35 degrees (see further explanations in the text). For display purposes we used tick marks 
 separated by  5$"$. The solid horizontal line in the $A_{\rm v}$ maps represent 5\,kpc.}
\label{teaser}
\end{figure}
\end{landscape}

The resolved synthesis (beyond the inner $\sim$ 1.5'') will be discussed in Paper II, and preliminary results are shown in Fig. \ref{teaser} for two sets of AGN and control sample galaxies: the first set corresponds to an early-type host and the second to a late-type host. The area covered by the central
fiber of the MaNGA array, at the mean distance of these galaxies, is around 1.5\,kpc.
In these figures, we have further collapsed the ages in three bins: young ($x_y$, $t<10$\,Myr), intermediate age ($x_i$, 51\,Myr$\le t \le$2\,Gyr) and old ($x_o$, $t>2$\,Gyr). 
In the case of the early-type galaxy, it can be seen that the AGN host has: (1) a stronger reddening over the whole galaxy than the two control galaxies, suggesting the presence of more gas in the AGN host galaxy; (2) a larger contribution of the young component; (3) a larger contribution of the intermediate age component in the inner region and then it becomes similar to that of the second control galaxy (green) outwards; (4) a smaller contribution of the old component everywhere in the galaxy, being approximately the same as that of the second control galaxy in the outer part of the galaxy; (5) an average age that is smaller than those of the two controls in the inner region and similar to that of the second control in the outer parts of the galaxy.

In the case of the late-type galaxy, the AGN host has: (1) a similar reddening to that of the second control galaxy, being smaller than that of the first control galaxy; (2) a lower contribution of the young component than the first control everywhere, a lower contribution than the second control in the inner parts but similar outwards; (3) a smaller contribution of the intermediate-age component to that of the controls everywhere; (4) a larger contribution of the old component everywhere and a (5) a larger mean age everywhere.

In summary, we have find a difference in the resolved stellar population properties between the AGN and control galaxies for the early-type AGN host and found no difference in the late-type AGN host. We note that the difference between the stellar population of the early-type AGN host and those of the control galaxies is not restricted to the nucleus, but is observed out to about 0.6 effective radii. In Paper II we will present the results for the whole sample, as well as a statistical analysis to investigate the difference between the AGN and control sample in terms of their resolved stellar population properties. 

\section{Conclusions}

We have characterized the first 62 AGN host galaxies observed in the MaNGA survey, defining and also characterizing a control sample of 2 galaxies for each AGN, matched according to global properties of the host galaxy such as stellar mass, distance, inclination and galaxy type. We compare the stellar population properties of the two samples within the inner $\sim$ 1.5'' radius around the nucleus (1--3\,kpc at their typical distances) using stellar population synthesis of their   SDSS-III spectra and the central spectrum of the MaNGA array, which are identical within the uncertainties. This study will be followed by the study of the resolved stellar population (Paper II) and gas properties (Paper III) over the whole region of the galaxy covered by the MaNGA observations.

The main results of this paper are:

\begin{itemize}
\item The stellar mass, redshift, $r$-band absolute magnitude, concentration and asymmetry distributions of the AGN hosts are well matched by those of the control sample. The galaxy morphologies are similar, with $\approx$65 per cent spiral and $\approx$33 per cent spheroidal galaxies. The luminosity $L(\rm{[OIII]})$, however, is markedly lower ($\sim$1.4\,dex) for the control sample galaxies, as expected. 

\item Only 17 AGN of our sample have AGN luminosities larger than $L(\rm{[OIII]})\geq 3.8\times 10^{40}$\,erg\,s$^{-1}$ (which we call strong AGN). Of these, three show signs of disturbed morphology, a much higher proportion than among the remaining, weak AGN.

\item The stellar population of the 45 weak AGN ($L(\rm{[OIII]})$ < $3.8\times 10^{40}$ erg s$^{-1}$) is dominated (down  to 60\% contribution to the light at 5700\AA) by old stars (with ages $4.0\leq t \leq 13.1\,\times 10^9$ yr) with a smaller contribution (up to 40\%) of intermediate age stars ($ 0.64 \leq t \leq 2.5\,\times 10^9$ yr) and almost no contribution of younger stars; similar results are observed for the control sample;

\item The strong AGN, on the other hand, show, on average, a larger contribution (of up to $\approx$ 20\%)  of younger ($\le 5\,\times 10^7$ yr) stars and a decreased contribution of the old stars relative to both the weak AGN and control sample;

\item A correlation between the stellar population properties and the AGN luminosity, extending also to lower L[OIII] values, is found when  pairing each AGN with its controls, via differential properties $AGN-control$, evidencing the importance in a careful selection of a control sample;

\item Via this pairing, we find a correlation between the diferential contribution (AGN - control) of the stellar population age component $x_{yo}$ ($25-48\times 10^6$yr) and the AGN luminosity L[OIII], and an inverse correlation between the differential contribution of the old age component $x_o$ ($4-13\times 10^9$yr) and L[OIII];

\item The pairing also reveals a trend in the mean differential age (AGN-control) and the AGN luminosity, in the sense that more luminous AGN are younger than the control sample. There is also a trend for weak AGN to be older than the control sample, while those with intermediate luminosity show similar ages to those of the control galaxies.

\end{itemize}

In summary, our results point to a difference between the stellar population of the AGN hosts and control sample that is correlated with the AGN luminosity: the most luminous the AGN, the younger is its stellar population in the inner kiloparsecs. This result supports the evolutionary scenario \citep{sb01,davies07} in which episodes of nuclear activity are preceded by episodes of star formation in the galaxy as both star formation and the nuclear activity feed on gas accretion towards the central regions of the galaxy.

In forthcoming papers we will expand the present investigation to include more AGN as they become observed with MaNGA, studying also the spatially resolved properties of the stellar population (Paper II), the gas excitation (Paper III) and kinematics, as well as the environment of our AGN sample as compared with those of the control sample using the spatially-resolved spectroscopic data from MaNGA.

\section*{Acknowledgements}

We would like to thank the support of the Instituto Nacional de Ci\^encia
e Tecnologia (INCT) e-Universe project (CNPq grant 465376/2014-2).
R.A.R. acknowledges support from FAPERGS (project no. 2366-2551/14-0) and CNPq (project no. 470090/2013-8 and 302683/2013-5).
R.R. thanks to FAPERGS (16/2551-0000251-7) and CNPq for financial support.

Funding for the Sloan Digital Sky Survey IV has been provided by the Alfred P. Sloan Foundation and the Participating Institutions.  SDSS-IV acknowledges
support and resources from the Center for High-Performance Computing at the University of Utah. The SDSS web site is www.sdss.org. SDSS-IV is managed by the
Astrophysical Research Consortium for the Participating Institutions of the SDSS Collaboration including the Brazilian Participation Group, the Carnegie Institution for Science,
Carnegie  Mellon  University,  the  Chilean  Participation  Group,  Harvard-Smithsonian Center for Astrophysics, Instituto de Astrof\'isica de Canarias, The Johns Hopkins University, Kavli Institute for the Physics and Mathematics of the Universe (IPMU) / University of Tokyo, Lawrence Berkeley National Laboratory, Leibniz Institut f\"ur Astro
physik  Potsdam  (AIP),  Max-Planck-Institut  f\"ur  Astrophysik  (MPA Garching),  Max-Planck-Institut f\"ur Extraterrestrische Physik (MPE), Max-Planck-Institut f\"ur Astronomie
(MPIA Heidelberg), National Astronomical Observatory of China, New Mexico State University, New York University, The Ohio State University, Pennsylvania State University, 
Shanghai  Astronomical  Observatory,  United  Kingdom  Participation  Group, Universidad Nacional Aut\'onoma de M\'exico, University of Arizona, University of Colorado Boulder,
University of Portsmouth, University of Utah, University of Washington, University of Wisconsin, Vanderbilt University, and Yale University.








\appendix

\section{SDSS-III combined \lowercase{\textit{ugriz}} negative images of MaNGA MPL-5 AGN and control sample objects}

\section{Detailed results from the stellar population synthesis}

\begin{landscape}
\begin{table}
\caption{Synthesis results for AGN in MaNGA MPL5. (1) galaxy identification in the MaNGA survey; (2) flux contribution of the featureless continuum; (3)-(8): percentual of the flux corresponding to each of the six age bins defined in Sect. 3.1; (9)-(14): mass contribution of the same age bins; (15) visual extinction in magnitudes; (16)-(17): light and mass weighted mean stellar age; (18)-(19): light and mass weighted mean stellar metallicity; (20): mean relative deviation between the fit and the galaxy spectrum.}
\label{synthesisagn}
\begin{tabular}{ccccccccccccccccccccc}
\hline
mangaID & FC  & xyy & xyo & xIy & xII & xIo & xo  &  myy & myo  & mIy  & mII  & mIo  & mo   & $A_V$ & $\langle t_L\rangle$ & $\langle t_M\rangle$ & $\langle Z_L\rangle$ & $\langle Z_M\rangle$ & Adev  \\
 (1)    & (2) & (3) & (4) & (5) & (6) & (7) & (8) & (9)  & (10) & (11) & (12) & (13) & (14) & (15)  & (16)    & (17)    & (18)    & (19)    & (20)  \\
\hline
1-109056 &  3.38 & 4.25 & 0.00 & 0.00 & 13.65 & 0.00 & 76.37 & 0.07 & 0.00 & 0.00 & 2.54 & 0.00 & 97.39 & 0.14 & 9.75 & 10.07 & 0.01957 & 0.01864 & 2.90 \\
1-121532 &  4.42 & 4.63 & 1.96 & 0.00 & 17.07 & 0.00 & 71.26 & 0.08 & 0.06 & 0.00 & 2.85 & 0.00 & 97.01 & 0.50 & 9.70 & 10.08 & 0.01965 & 0.02309 & 5.40 \\
1-135044 &  7.63 & 0.00 & 0.00 & 0.00 & 0.00 & 4.03 & 85.79 & 0.00 & 0.00 & 0.00 & 0.00 & 1.22 & 98.78 & 0.14 & 9.92 & 10.02 & 0.01727 & 0.01877 & 2.58 \\
1-135285 &  0.04 & 0.00 & 0.00 & 0.00 & 0.00 & 0.00 & 97.15 & 0.00 & 0.00 & 0.00 & 0.00 & 0.00 & 100.00 & 0.17 & 10.00 & 10.07 & 0.01475 & 0.01761 & 2.76 \\
1-135641 &  4.24 & 2.54 & 0.00 & 0.00 & 9.54 & 0.00 & 80.90 & 0.04 & 0.00 & 0.00 & 1.34 & 0.00 & 98.62 & 0.91 & 9.88 & 10.10 & 0.01857 & 0.02283 & 2.45 \smallskip \\
1-137883 &  0.00 & 10.51 & 9.50 & 0.00 & 2.28 & 40.64 & 34.75 & 0.16 & 0.31 & 0.00 & 0.62 & 14.82 & 84.10 & 1.58 & 9.12 & 9.98 & 0.01446 & 0.02749 & 3.28 \\
1-148068 &  3.61 & 0.00 & 0.00 & 0.00 & 0.00 & 13.03 & 79.76 & 0.00 & 0.00 & 0.00 & 0.00 & 3.14 & 96.86 & 0.26 & 9.98 & 10.08 & 0.0246 & 0.02462 & 3.86 \\
1-149211 &  12.13 & 0.00 & 0.00 & 0.00 & 0.00 & 49.60 & 36.05 & 0.00 & 0.00 & 0.00 & 0.00 & 27.91 & 72.09 & 0.18 & 9.55 & 9.83 & 0.03361 & 0.02262 & 3.96 \\
1-163831 &  3.93 & 1.36 & 0.00 & 0.00 & 0.00 & 9.75 & 81.93 & 0.02 & 0.00 & 0.00 & 0.00 & 2.32 & 97.66 & 0.24 & 9.90 & 10.06 & 0.01905 & 0.02243 & 3.05 \\
1-166919 &  4.98 & 1.72 & 0.00 & 0.00 & 4.70 & 13.33 & 72.58 & 0.03 & 0.00 & 0.00 & 0.83 & 2.28 & 96.86 & 0.23 & 9.84 & 10.08 & 0.01942 & 0.01996 & 2.39 \smallskip \\
1-167688 &  3.56 & 3.80 & 1.15 & 0.00 & 34.40 & 25.77 & 30.48 & 0.12 & 0.05 & 0.00 & 11.29 & 10.72 & 77.83 & 0.15 & 9.27 & 9.87 & 0.01905 & 0.02004 & 2.16 \\
1-173958 &  7.42 & 16.37 & 0.85 & 0.00 & 19.93 & 20.47 & 33.50 & 0.28 & 0.02 & 0.00 & 6.16 & 8.44 & 85.11 & 0.62 & 9.05 & 9.96 & 0.02302 & 0.02879 & 3.47 \\
1-198153 &  0.00 & 1.69 & 0.00 & 0.00 & 0.00 & 0.00 & 96.20 & 0.02 & 0.00 & 0.00 & 0.00 & 0.00 & 99.98 & 0.41 & 10.04 & 10.11 & 0.01435 & 0.01666 & 2.63 \\
1-198182 &  4.95 & 0.00 & 0.00 & 0.00 & 0.00 & 0.00 & 92.01 & 0.00 & 0.00 & 0.00 & 0.00 & 0.00 & 100.00 & -0.06 & 10.11 & 10.11 & 0.02216 & 0.02484 & 1.87 \\
1-201561 &  4.08 & 0.00 & 0.00 & 0.00 & 0.00 & 0.00 & 91.27 & 0.00 & 0.00 & 0.00 & 0.00 & 0.00 & 100.00 & -0.03 & 10.11 & 10.11 & 0.02364 & 0.02726 & 3.79 \smallskip \\
1-209980 &  17.08 & 0.00 & 0.00 & 0.00 & 0.00 & 0.00 & 81.52 & 0.00 & 0.00 & 0.00 & 0.00 & 0.00 & 100.00 & 0.03 & 10.11 & 10.11 & 0.01671 & 0.01897 & 2.00 \\
1-210646 &  3.37 & 4.78 & 0.00 & 0.00 & 0.00 & 18.12 & 71.64 & 0.11 & 0.00 & 0.00 & 0.00 & 7.47 & 92.42 & 0.45 & 9.54 & 9.92 & 0.01784 & 0.01777 & 4.77 \\
1-211311 &  4.34 & 0.00 & 0.00 & 0.00 & 0.00 & 6.11 & 85.90 & 0.00 & 0.00 & 0.00 & 0.00 & 1.67 & 98.33 & -0.02 & 9.97 & 10.05 & 0.0176 & 0.01795 & 2.45 \\
1-217050 &  1.26 & 1.12 & 0.00 & 0.00 & 0.00 & 0.00 & 94.71 & 0.01 & 0.00 & 0.00 & 0.00 & 0.00 & 99.99 & 0.00 & 10.07 & 10.11 & 0.02053 & 0.0238 & 1.64 \\
1-22301 &  7.65 & 2.15 & 0.00 & 0.00 & 0.00 & 11.75 & 74.97 & 0.02 & 0.00 & 0.00 & 0.00 & 2.40 & 97.58 & 0.18 & 9.89 & 10.08 & 0.01894 & 0.0237 & 2.59 \smallskip \\
1-229010 &  2.42 & 0.00 & 0.00 & 0.00 & 0.00 & 0.00 & 93.18 & 0.00 & 0.00 & 0.00 & 0.00 & 0.00 & 100.00 & -0.10 & 10.06 & 10.09 & 0.01987 & 0.02236 & 1.58 \\
1-234618 &  0.00 & 3.91 & 0.00 & 0.00 & 2.10 & 2.29 & 88.13 & 0.06 & 0.00 & 0.00 & 0.42 & 0.61 & 98.92 & 0.78 & 9.78 & 10.04 & 0.01416 & 0.01897 & 5.22 \\
1-23979 &  1.60 & 5.66 & 0.00 & 0.00 & 25.10 & 0.00 & 65.77 & 0.10 & 0.00 & 0.00 & 5.11 & 0.00 & 94.79 & 0.54 & 9.57 & 10.05 & 0.02522 & 0.02186 & 2.63 \\
1-24148 &  8.51 & 0.00 & 0.00 & 0.00 & 0.00 & 5.64 & 83.67 & 0.00 & 0.00 & 0.00 & 0.00 & 1.32 & 98.68 & 0.26 & 10.05 & 10.10 & 0.02232 & 0.02433 & 2.28 \\
1-248389 &  0.00 & 3.03 & 0.00 & 5.12 & 11.80 & 38.14 & 39.81 & 0.07 & 0.00 & 0.51 & 2.83 & 9.95 & 86.65 & 0.45 & 9.41 & 9.98 & 0.01949 & 0.03118 & 1.99 \smallskip \\
1-248420 &  9.48 & 0.00 & 0.00 & 0.00 & 0.00 & 0.00 & 87.58 & 0.00 & 0.00 & 0.00 & 0.00 & 0.00 & 100.00 & 0.18 & 10.09 & 10.09 & 0.01447 & 0.01568 & 3.07 \\
1-25554 &  0.00 & 3.36 & 0.00 & 0.00 & 0.00 & 34.06 & 59.68 & 0.06 & 0.00 & 0.00 & 0.00 & 9.95 & 89.99 & 0.32 & 9.68 & 10.03 & 0.01957 & 0.01902 & 2.66 \\
1-256446 &  1.36 & 0.00 & 0.00 & 0.00 & 0.00 & 0.00 & 94.32 & 0.00 & 0.00 & 0.00 & 0.00 & 0.00 & 100.00 & -0.04 & 10.11 & 10.11 & 0.01799 & 0.01905 & 2.96 \\
1-25725 &  11.87 & 1.28 & 0.00 & 0.00 & 0.00 & 0.00 & 85.18 & 0.00 & 0.00 & 0.00 & 0.00 & 0.00 & 100.00 & 0.17 & 10.06 & 10.11 & 0.03177 & 0.03529 & 3.20 \\
1-258599 &  9.11 & 13.62 & 6.83 & 2.11 & 14.44 & 12.09 & 39.36 & 0.20 & 0.22 & 0.29 & 5.15 & 5.75 & 88.39 & 0.62 & 9.01 & 9.92 & 0.02181 & 0.03697 & 3.19 \smallskip \\
1-258774 &  1.79 & 4.81 & 0.00 & 0.00 & 10.70 & 27.62 & 51.72 & 0.04 & 0.00 & 0.00 & 2.43 & 8.96 & 88.57 & 0.41 & 9.56 & 10.00 & 0.02701 & 0.0221 & 2.02 \\
1-259142 &  3.62 & 0.00 & 0.00 & 0.00 & 0.00 & 0.00 & 92.92 & 0.00 & 0.00 & 0.00 & 0.00 & 0.00 & 100.00 & -0.00 & 10.11 & 10.11 & 0.02069 & 0.02218 & 2.20 \\
1-269632 &  4.49 & 18.98 & 8.39 & 0.00 & 37.97 & 18.93 & 9.14 & 0.51 & 0.50 & 0.00 & 24.60 & 16.72 & 57.68 & 0.65 & 8.61 & 9.65 & 0.02923 & 0.04336 & 3.64 \\
1-277552 &  0.00 & 5.96 & 0.00 & 0.00 & 10.34 & 0.00 & 79.53 & 0.10 & 0.00 & 0.00 & 1.55 & 0.00 & 98.35 & 0.64 & 9.71 & 10.04 & 0.01871 & 0.0161 & 4.72 \\
1-279073 &  5.45 & 0.00 & 0.00 & 0.00 & 0.00 & 0.00 & 90.99 & 0.00 & 0.00 & 0.00 & 0.00 & 0.00 & 100.00 & -0.13 & 10.11 & 10.11 & 0.02647 & 0.02996 & 2.19 \smallskip \\
1-279147 &  5.08 & 5.23 & 0.00 & 0.00 & 18.03 & 37.63 & 31.30 & 0.09 & 0.00 & 0.00 & 6.78 & 20.13 & 73.00 & 0.56 & 9.25 & 9.81 & 0.03086 & 0.0274 & 2.75 \\
1-279666 &  0.00 & 3.12 & 0.00 & 0.00 & 7.44 & 36.36 & 49.72 & 0.07 & 0.00 & 0.00 & 1.90 & 11.48 & 86.55 & 0.26 & 9.54 & 9.93 & 0.01997 & 0.02375 & 2.44 \\
1-279676 &  0.00 & 1.99 & 0.00 & 0.00 & 7.49 & 19.39 & 69.58 & 0.03 & 0.00 & 0.00 & 1.48 & 5.64 & 92.84 & 0.32 & 9.75 & 10.04 & 0.02402 & 0.01853 & 3.36 \\
1-321739 &  0.00 & 2.74 & 2.19 & 0.00 & 15.14 & 9.42 & 68.96 & 0.03 & 0.05 & 0.00 & 2.80 & 2.45 & 94.67 & 1.08 & 9.69 & 10.06 & 0.02029 & 0.02271 & 2.96 \\
1-338922 &  0.00 & 14.28 & 0.00 & 0.00 & 0.00 & 32.92 & 51.74 & 0.16 & 0.00 & 0.00 & 0.00 & 17.42 & 82.41 & 0.48 & 9.33 & 9.92 & 0.01828 & 0.02573 & 6.43 \\
\hline
\end{tabular}
\end{table}
\end{landscape}

\begin{landscape}
\begin{table}
\contcaption{}
\begin{tabular}{ccccccccccccccccccccc}
\hline
mangaID & FC  & xyy & xyo & xIy & xII & xIo & xo  &  myy & myo  & mIy  & mII  & mIo  & mo   & $A_V$ & $\langle t_L\rangle$ & $\langle t_M\rangle$ & $\langle Z_L\rangle$ & $\langle Z_M\rangle$ & Adev  \\
 (1)    & (2) & (3) & (4) & (5) & (6) & (7) & (8) & (9)  & (10) & (11) & (12) & (13) & (14) & (15)  & (16)    & (17)    & (18)    & (19)    & (20)  \\
\hline
1-339094 &  7.03 & 0.00 & 0.00 & 0.00 & 7.66 & 23.20 & 59.24 & 0.00 & 0.00 & 0.00 & 1.90 & 8.37 & 89.73 & 0.29 & 9.72 & 9.98 & 0.02492 & 0.01955 & 1.86 \\
1-339163 &  5.93 & 0.00 & 0.00 & 0.00 & 0.00 & 0.00 & 92.23 & 0.00 & 0.00 & 0.00 & 0.00 & 0.00 & 100.00 & 0.00 & 10.08 & 10.10 & 0.01674 & 0.02013 & 1.94 \\
1-351790 &  5.46 & 2.08 & 2.60 & 0.00 & 43.77 & 30.59 & 13.55 & 0.10 & 0.16 & 0.00 & 22.41 & 22.30 & 55.03 & 0.41 & 9.16 & 9.68 & 0.02216 & 0.02057 & 2.20 \\
1-37036 &  0.00 & 1.59 & 0.00 & 0.00 & 0.00 & 2.20 & 93.97 & 0.02 & 0.00 & 0.00 & 0.00 & 0.27 & 99.71 & 0.22 & 10.03 & 10.11 & 0.01886 & 0.02278 & 2.11 \\
1-373161 &  4.82 & 0.00 & 0.00 & 0.00 & 0.00 & 0.00 & 91.52 & 0.00 & 0.00 & 0.00 & 0.00 & 0.00 & 100.00 & 0.16 & 10.11 & 10.11 & 0.02199 & 0.02394 & 2.64 \smallskip \\
1-44303 &  11.93 & 1.62 & 0.00 & 0.00 & 4.50 & 12.94 & 68.16 & 0.03 & 0.00 & 0.00 & 1.04 & 7.58 & 91.34 & 0.35 & 9.82 & 10.02 & 0.01984 & 0.0175 & 3.13 \\
1-44379 &  12.87 & 0.00 & 0.00 & 0.00 & 0.92 & 5.16 & 78.93 & 0.00 & 0.00 & 0.00 & 0.18 & 0.87 & 98.95 & 0.17 & 10.04 & 10.10 & 0.0124 & 0.01433 & 2.08 \\
1-460812 &  1.82 & 1.94 & 0.00 & 0.00 & 0.00 & 3.39 & 89.71 & 0.02 & 0.00 & 0.00 & 0.00 & 0.78 & 99.20 & 0.73 & 9.92 & 10.08 & 0.01719 & 0.0231 & 3.26 \\
1-48116 &  0.00 & 6.62 & 0.00 & 0.00 & 17.37 & 11.44 & 62.82 & 0.12 & 0.00 & 0.00 & 3.80 & 3.19 & 92.89 & 0.56 & 9.54 & 10.04 & 0.0195 & 0.01876 & 1.92 \\
1-491229 &  6.52 & 0.00 & 0.00 & 0.00 & 0.00 & 0.00 & 91.23 & 0.00 & 0.00 & 0.00 & 0.00 & 0.00 & 100.00 & 0.00 & 10.11 & 10.11 & 0.02215 & 0.02543 & 2.03 \smallskip \\
1-519742 &  2.59 & 0.00 & 0.00 & 0.00 & 0.00 & 71.52 & 23.59 & 0.00 & 0.00 & 0.00 & 0.00 & 64.99 & 35.01 & 0.26 & 9.32 & 9.37 & 0.033 & 0.02945 & 5.97 \\
1-542318 &  2.68 & 0.00 & 0.00 & 0.00 & 0.00 & 12.83 & 82.73 & 0.00 & 0.00 & 0.00 & 0.00 & 4.09 & 95.91 & 0.30 & 9.99 & 10.08 & 0.01777 & 0.01604 & 4.66 \\
1-558912 &  11.71 & 0.00 & 0.00 & 0.00 & 0.00 & 12.28 & 72.81 & 0.00 & 0.00 & 0.00 & 0.00 & 2.94 & 97.06 & 0.03 & 9.98 & 10.09 & 0.02779 & 0.02649 & 3.45 \\
1-604761 &  2.77 & 0.00 & 0.00 & 0.00 & 0.00 & 0.00 & 93.55 & 0.00 & 0.00 & 0.00 & 0.00 & 0.00 & 100.00 & 0.11 & 10.11 & 10.11 & 0.01829 & 0.02073 & 2.67 \\
1-72322 &  16.54 & 0.00 & 0.00 & 0.00 & 0.00 & 0.00 & 80.48 & 0.00 & 0.00 & 0.00 & 0.00 & 0.00 & 100.00 & 0.19 & 10.11 & 10.11 & 0.02889 & 0.03089 & 3.34 \smallskip \\
1-91016 &  0.00 & 4.63 & 0.00 & 0.00 & 0.00 & 56.04 & 37.49 & 0.11 & 0.00 & 0.00 & 0.00 & 29.37 & 70.52 & 0.70 & 9.37 & 9.75 & 0.01953 & 0.02314 & 4.36 \\
1-92866 &  0.00 & 1.29 & 0.00 & 0.00 & 0.00 & 0.00 & 95.45 & 0.02 & 0.00 & 0.00 & 0.00 & 0.00 & 99.98 & 0.32 & 10.02 & 10.10 & 0.01964 & 0.02229 & 2.97 \\
1-94604 &  9.35 & 0.00 & 0.00 & 0.00 & 0.00 & 13.18 & 74.83 & 0.00 & 0.00 & 0.00 & 0.00 & 7.74 & 92.26 & -0.01 & 9.82 & 9.97 & 0.02021 & 0.02463 & 3.16 \\
1-94784 &  2.90 & 0.00 & 0.00 & 0.00 & 0.00 & 36.69 & 55.88 & 0.00 & 0.00 & 0.00 & 0.00 & 12.99 & 87.01 & 0.24 & 9.68 & 9.95 & 0.02994 & 0.02287 & 2.13 \\
1-95092 &  0.00 & 3.74 & 0.00 & 0.00 & 11.56 & 0.00 & 82.36 & 0.03 & 0.00 & 0.00 & 2.06 & 0.00 & 97.91 & 0.39 & 9.75 & 10.06 & 0.02084 & 0.0195 & 2.03 \smallskip \\
1-95585 &  4.16 & 0.00 & 0.00 & 0.00 & 0.00 & 0.63 & 91.02 & 0.00 & 0.00 & 0.00 & 0.00 & 0.22 & 99.78 & -0.02 & 10.11 & 10.11 & 0.02265 & 0.02563 & 3.11 \\
1-96075 &  2.81 & 6.57 & 0.00 & 0.00 & 0.00 & 16.27 & 71.05 & 0.06 & 0.00 & 0.00 & 0.00 & 5.30 & 94.63 & 0.38 & 9.63 & 10.01 & 0.01634 & 0.01836 & 4.50 \\
\hline
\end{tabular}
\end{table}
\end{landscape}

\begin{landscape}
\begin{table}
\caption{Synthesis results for the control sample. The columns are the same as in Table~\ref{synthesisagn}.}
\label{synthesiscontrol}
\begin{tabular}{ccccccccccccccccccccc}
\hline
mangaID & FC  & xyy & xyo & xIy & xII & xIo & xo  &  myy & myo  & mIy  & mII  & mIo  & mo   & $A_V$ & $\langle t_L\rangle$ & $\langle t_M\rangle$ & $\langle Z_L\rangle$ & $\langle Z_M\rangle$ & Adev  \\
 (1)    & (2) & (3) & (4) & (5) & (6) & (7) & (8) & (9)  & (10) & (11) & (12) & (13) & (14) & (15)  & (16)    & (17)    & (18)    & (19)    & (20)  \\
\hline
1-109493 &  2.97 & 0.00 & 0.00 & 0.00 & 0.00 & 0.00 & 91.99 & 0.00 & 0.00 & 0.00 & 0.00 & 0.00 & 100.00 & -0.15 & 10.11 & 10.11 & 0.0219 & 0.02306 & 3.52 \\
1-114306 &  1.09 & 8.07 & 0.00 & 0.00 & 7.78 & 34.00 & 46.61 & 0.13 & 0.00 & 0.00 & 2.22 & 12.92 & 84.73 & 0.44 & 9.39 & 9.95 & 0.01419 & 0.0174 & 4.47 \\
1-121717 &  0.00 & 4.42 & 0.00 & 0.00 & 4.38 & 32.26 & 56.85 & 0.07 & 0.00 & 0.00 & 0.88 & 8.77 & 90.28 & 0.53 & 9.63 & 10.03 & 0.01883 & 0.02178 & 3.11 \\
1-134239 &  0.00 & 0.00 & 0.00 & 0.00 & 0.00 & 14.05 & 82.24 & 0.00 & 0.00 & 0.00 & 0.00 & 3.52 & 96.48 & 0.11 & 9.91 & 10.06 & 0.02297 & 0.02196 & 3.03 \\
1-135371 &  0.00 & 2.66 & 0.00 & 0.00 & 14.05 & 0.00 & 82.30 & 0.02 & 0.00 & 0.00 & 2.41 & 0.00 & 97.57 & 0.65 & 9.84 & 10.09 & 0.02158 & 0.02076 & 2.76 \smallskip \\
1-135372 &  0.00 & 0.00 & 3.19 & 0.00 & 0.00 & 0.00 & 96.37 & 0.00 & 0.05 & 0.00 & 0.00 & 0.00 & 99.95 & 0.02 & 10.03 & 10.11 & 0.01941 & 0.0215 & 1.69 \\
1-135502 &  0.60 & 0.00 & 0.00 & 0.00 & 0.00 & 0.00 & 94.52 & 0.00 & 0.00 & 0.00 & 0.00 & 0.00 & 100.00 & 0.13 & 10.03 & 10.08 & 0.01767 & 0.02035 & 2.18 \\
1-135625 &  12.87 & 2.50 & 12.97 & 0.00 & 33.15 & 18.44 & 17.30 & 0.10 & 0.60 & 0.00 & 15.10 & 11.27 & 72.93 & 0.35 & 8.97 & 9.82 & 0.03015 & 0.03899 & 1.75 \\
1-135810 &  0.00 & 0.00 & 0.00 & 0.00 & 0.00 & 41.05 & 57.21 & 0.00 & 0.00 & 0.00 & 0.00 & 16.27 & 83.73 & 0.30 & 9.61 & 9.89 & 0.02852 & 0.02237 & 2.64 \\
1-136125 &  0.00 & 0.00 & 0.00 & 0.00 & 0.00 & 36.16 & 60.24 & 0.00 & 0.00 & 0.00 & 0.00 & 13.23 & 86.77 & 0.29 & 9.66 & 9.93 & 0.02761 & 0.0219 & 4.52 \smallskip \\
1-166691 &  0.59 & 1.18 & 0.00 & 0.00 & 0.00 & 7.98 & 86.24 & 0.01 & 0.00 & 0.00 & 0.00 & 1.53 & 98.46 & -0.14 & 10.00 & 10.10 & 0.02201 & 0.02557 & 3.00 \\
1-166947 &  3.68 & 0.00 & 0.00 & 0.00 & 0.00 & 0.00 & 92.37 & 0.00 & 0.00 & 0.00 & 0.00 & 0.00 & 100.00 & -0.10 & 10.11 & 10.11 & 0.02198 & 0.02528 & 3.27 \\
1-167334 &  0.00 & 3.03 & 0.00 & 0.00 & 56.89 & 36.57 & 1.18 & 0.11 & 0.00 & 0.00 & 49.04 & 42.99 & 7.86 & 0.55 & 9.04 & 9.20 & 0.03621 & 0.03418 & 1.65 \\
1-177493 &  3.34 & 0.00 & 0.00 & 0.00 & 0.00 & 19.37 & 74.21 & 0.00 & 0.00 & 0.00 & 0.00 & 5.77 & 94.23 & 0.03 & 9.82 & 10.02 & 0.02338 & 0.02186 & 2.30 \\
1-178838 &  12.76 & 0.00 & 0.00 & 0.00 & 3.31 & 17.56 & 62.46 & 0.00 & 0.00 & 0.00 & 0.62 & 4.47 & 94.91 & -0.12 & 9.91 & 10.07 & 0.02277 & 0.02485 & 1.94 \smallskip \\
1-210173 &  5.84 & 0.00 & 0.00 & 0.00 & 2.29 & 31.23 & 55.48 & 0.00 & 0.00 & 0.00 & 0.33 & 10.10 & 89.57 & 0.24 & 9.74 & 10.00 & 0.0307 & 0.02489 & 4.72 \\
1-210593 &  0.00 & 0.00 & 0.00 & 0.00 & 6.32 & 0.00 & 91.27 & 0.00 & 0.00 & 0.00 & 1.07 & 0.00 & 98.93 & 0.28 & 9.91 & 10.05 & 0.0209 & 0.02503 & 3.43 \\
1-210614 &  0.00 & 1.24 & 0.00 & 0.00 & 0.00 & 0.00 & 94.97 & 0.01 & 0.00 & 0.00 & 0.00 & 0.00 & 99.98 & 0.09 & 10.06 & 10.11 & 0.02004 & 0.02273 & 2.48 \\
1-210700 &  0.00 & 0.00 & 0.00 & 0.00 & 0.00 & 0.00 & 96.81 & 0.00 & 0.00 & 0.00 & 0.00 & 0.00 & 100.00 & 0.16 & 10.09 & 10.11 & 0.02369 & 0.02721 & 5.73 \\
1-210784 &  0.00 & 2.87 & 0.00 & 0.00 & 1.19 & 4.69 & 88.16 & 0.03 & 0.00 & 0.00 & 0.11 & 0.56 & 99.29 & 0.11 & 9.93 & 10.11 & 0.02379 & 0.02871 & 2.34 \smallskip \\
1-210962 &  0.00 & 1.41 & 0.00 & 0.00 & 5.66 & 8.40 & 81.53 & 0.02 & 0.00 & 0.00 & 0.89 & 1.76 & 97.33 & 0.15 & 9.92 & 10.09 & 0.02033 & 0.02174 & 1.64 \\
1-211063 &  0.00 & 1.39 & 0.00 & 0.00 & 0.00 & 18.78 & 75.78 & 0.02 & 0.00 & 0.00 & 0.00 & 4.88 & 95.10 & 0.28 & 9.87 & 10.07 & 0.02286 & 0.0192 & 2.32 \\
1-211074 &  0.00 & 1.86 & 0.00 & 0.00 & 0.00 & 0.00 & 94.99 & 0.02 & 0.00 & 0.00 & 0.00 & 0.00 & 99.97 & 0.13 & 10.03 & 10.11 & 0.01627 & 0.01925 & 2.31 \\
1-211079 &  0.00 & 0.00 & 0.00 & 0.00 & 5.97 & 0.00 & 92.39 & 0.00 & 0.00 & 0.00 & 0.59 & 0.00 & 99.41 & 0.03 & 10.03 & 10.10 & 0.02051 & 0.02551 & 1.93 \\
1-211082 &  0.00 & 2.26 & 0.00 & 0.00 & 7.17 & 0.00 & 88.14 & 0.02 & 0.00 & 0.00 & 0.99 & 0.00 & 98.98 & 0.23 & 9.93 & 10.10 & 0.02593 & 0.02656 & 2.23 \smallskip \\
1-211100 &  0.00 & 1.46 & 0.00 & 0.00 & 0.00 & 0.00 & 95.79 & 0.02 & 0.00 & 0.00 & 0.00 & 0.00 & 99.98 & 0.04 & 10.05 & 10.11 & 0.01701 & 0.02087 & 1.85 \\
12-129446 &  1.31 & 4.33 & 0.00 & 0.00 & 0.01 & 40.22 & 51.60 & 0.08 & 0.00 & 0.00 & 0.00 & 12.01 & 87.91 & 0.33 & 9.59 & 10.01 & 0.02053 & 0.02209 & 2.84 \\
1-216958 &  15.73 & 0.00 & 0.00 & 0.00 & 13.14 & 11.64 & 56.77 & 0.00 & 0.00 & 0.00 & 2.41 & 2.68 & 94.91 & -0.08 & 9.77 & 10.05 & 0.02646 & 0.02707 & 1.60 \\
1-218280 &  2.44 & 0.00 & 0.00 & 0.00 & 0.00 & 0.00 & 94.10 & 0.00 & 0.00 & 0.00 & 0.00 & 0.00 & 100.00 & -0.14 & 10.11 & 10.11 & 0.0209 & 0.02264 & 3.90 \\
1-218427 &  2.55 & 0.00 & 0.00 & 0.00 & 0.00 & 0.00 & 94.12 & 0.00 & 0.00 & 0.00 & 0.00 & 0.00 & 100.00 & -0.11 & 10.11 & 10.11 & 0.02109 & 0.02295 & 3.88 \smallskip \\
1-235398 &  0.00 & 5.31 & 0.00 & 0.00 & 0.00 & 17.60 & 74.64 & 0.08 & 0.00 & 0.00 & 0.00 & 5.14 & 94.78 & 0.81 & 9.64 & 10.01 & 0.0122 & 0.01717 & 2.83 \\
1-235587 &  0.00 & 1.71 & 0.00 & 0.00 & 6.86 & 0.00 & 87.88 & 0.02 & 0.00 & 0.00 & 1.13 & 0.00 & 98.85 & -0.07 & 9.92 & 10.08 & 0.01878 & 0.01863 & 2.09 \\
1-236099 &  3.82 & 0.00 & 0.00 & 0.00 & 22.14 & 28.78 & 42.23 & 0.00 & 0.00 & 0.00 & 6.66 & 12.26 & 81.08 & 0.23 & 9.57 & 9.93 & 0.02743 & 0.01906 & 2.65 \\
1-94514 &  0.00 & 1.21 & 0.00 & 0.00 & 0.00 & 13.14 & 82.88 & 0.02 & 0.00 & 0.00 & 0.00 & 3.02 & 96.96 & 0.00 & 9.93 & 10.08 & 0.02384 & 0.02302 & 2.23 \\
1-23731 &  0.00 & 1.71 & 0.00 & 0.00 & 0.00 & 2.08 & 91.90 & 0.02 & 0.00 & 0.00 & 0.00 & 0.45 & 99.52 & 0.02 & 9.93 & 10.07 & 0.01792 & 0.02215 & 2.01 \\smallskip \\
1-24099 &  0.00 & 0.91 & 0.00 & 0.00 & 16.33 & 0.00 & 80.68 & 0.01 & 0.00 & 0.00 & 2.69 & 0.00 & 97.30 & 0.02 & 9.88 & 10.08 & 0.0235 & 0.02102 & 1.85 \\
1-24246 &  6.94 & 0.00 & 0.00 & 0.00 & 30.11 & 0.95 & 58.39 & 0.00 & 0.00 & 0.00 & 6.37 & 0.17 & 93.46 & 0.18 & 9.71 & 10.03 & 0.0303 & 0.02221 & 6.43 \\
1-24416 &  0.00 & 1.01 & 0.00 & 0.00 & 0.00 & 0.00 & 96.33 & 0.01 & 0.00 & 0.00 & 0.00 & 0.00 & 99.99 & 0.25 & 9.97 & 10.07 & 0.01625 & 0.01925 & 1.94 \\
\hline
\end{tabular}
\end{table}
\end{landscape}

\begin{landscape}
\begin{table}
\contcaption{}
\begin{tabular}{ccccccccccccccccccccc}
\hline
mangaID & FC  & xyy & xyo & xIy & xII & xIo & xo  &  myy & myo  & mIy  & mII  & mIo  & mo   & $A_V$ & $\langle t_L\rangle$ & $\langle t_M\rangle$ & $\langle Z_L\rangle$ & $\langle Z_M\rangle$ & Adev  \\
 (1)    & (2) & (3) & (4) & (5) & (6) & (7) & (8) & (9)  & (10) & (11) & (12) & (13) & (14) & (15)  & (16)    & (17)    & (18)    & (19)    & (20)  \\
\hline
1-245774 &  0.00 & 1.23 & 0.00 & 0.00 & 5.15 & 33.24 & 57.82 & 0.02 & 0.00 & 0.00 & 1.06 & 9.54 & 89.38 & 0.07 & 9.72 & 10.02 & 0.02176 & 0.02083 & 2.58 \\
1-247417 &  0.00 & 6.75 & 0.00 & 0.00 & 1.71 & 53.99 & 33.94 & 0.16 & 0.00 & 0.00 & 0.49 & 21.04 & 78.30 & 0.73 & 9.38 & 9.94 & 0.01549 & 0.02187 & 2.97 \\
1-247456 &  0.00 & 5.43 & 0.00 & 0.00 & 9.06 & 38.44 & 44.44 & 0.11 & 0.00 & 0.00 & 2.22 & 12.55 & 85.12 & 0.43 & 9.49 & 9.99 & 0.01634 & 0.02028 & 3.45 \\
1-251279 &  0.00 & 1.13 & 0.00 & 0.00 & 2.77 & 11.06 & 80.72 & 0.02 & 0.00 & 0.00 & 0.47 & 2.54 & 96.97 & 0.20 & 9.90 & 10.07 & 0.01819 & 0.02016 & 1.97 \\
1-251871 &  6.57 & 0.00 & 0.00 & 0.00 & 0.00 & 0.00 & 89.51 & 0.00 & 0.00 & 0.00 & 0.00 & 0.00 & 100.00 & 0.06 & 10.11 & 10.11 & 0.02168 & 0.02298 & 5.18 \smallskip \\
1-256185 &  2.28 & 0.00 & 0.00 & 0.00 & 0.00 & 0.00 & 93.72 & 0.00 & 0.00 & 0.00 & 0.00 & 0.00 & 100.00 & -0.05 & 10.07 & 10.10 & 0.02002 & 0.02211 & 1.72 \\
1-256465 &  0.00 & 0.00 & 0.00 & 0.00 & 0.00 & 26.71 & 69.99 & 0.00 & 0.00 & 0.00 & 0.00 & 7.31 & 92.69 & -0.10 & 9.85 & 10.04 & 0.02669 & 0.02267 & 2.58 \\
1-25680 &  0.00 & 1.79 & 0.00 & 0.00 & 0.00 & 0.00 & 94.25 & 0.02 & 0.00 & 0.00 & 0.00 & 0.00 & 99.98 & -0.13 & 10.01 & 10.09 & 0.02893 & 0.03344 & 2.76 \\
1-25688 &  0.00 & 6.03 & 0.00 & 0.00 & 0.00 & 10.33 & 80.42 & 0.08 & 0.00 & 0.00 & 0.00 & 4.89 & 95.03 & 0.33 & 9.66 & 10.01 & 0.01729 & 0.01877 & 4.50 \\
1-258455 &  0.00 & 1.27 & 0.00 & 0.00 & 0.00 & 0.00 & 95.81 & 0.02 & 0.00 & 0.00 & 0.00 & 0.00 & 99.98 & 0.26 & 9.98 & 10.08 & 0.0147 & 0.01823 & 2.76 \smallskip \\
1-259650 &  0.00 & 1.69 & 0.00 & 0.00 & 0.00 & 0.00 & 94.87 & 0.02 & 0.00 & 0.00 & 0.00 & 0.00 & 99.98 & 0.23 & 10.04 & 10.11 & 0.02089 & 0.02388 & 1.96 \\
1-264513 &  0.00 & 9.38 & 0.00 & 4.55 & 15.78 & 36.93 & 30.91 & 0.15 & 0.00 & 0.57 & 5.19 & 13.71 & 80.37 & 0.35 & 9.15 & 9.93 & 0.01918 & 0.0212 & 1.82 \\
1-270160 &  0.00 & 1.21 & 0.00 & 0.00 & 0.00 & 0.00 & 94.65 & 0.01 & 0.00 & 0.00 & 0.00 & 0.00 & 99.99 & 0.39 & 10.06 & 10.11 & 0.02402 & 0.02761 & 3.05 \\
1-274646 &  0.43 & 2.52 & 0.67 & 0.00 & 15.02 & 52.28 & 25.36 & 0.04 & 0.02 & 0.00 & 5.27 & 25.68 & 68.98 & 0.41 & 9.33 & 9.81 & 0.03476 & 0.02703 & 1.93 \\
1-274663 &  1.97 & 0.00 & 0.00 & 0.00 & 0.00 & 0.00 & 95.14 & 0.00 & 0.00 & 0.00 & 0.00 & 0.00 & 100.00 & 0.07 & 10.11 & 10.11 & 0.01707 & 0.01898 & 1.61 \smallskip \\
1-276679 &  15.39 & 4.90 & 0.00 & 0.00 & 22.42 & 42.50 & 13.72 & 0.08 & 0.00 & 0.00 & 14.85 & 38.57 & 46.50 & 0.32 & 9.14 & 9.48 & 0.02879 & 0.03368 & 5.12 \\
1-282144 &  13.94 & 8.05 & 0.00 & 0.00 & 8.22 & 57.94 & 10.42 & 0.36 & 0.00 & 0.00 & 4.98 & 53.94 & 40.72 & 0.69 & 9.06 & 9.48 & 0.01918 & 0.0306 & 6.20 \\
1-283246 &  0.00 & 0.00 & 0.00 & 0.00 & 0.00 & 31.06 & 64.84 & 0.00 & 0.00 & 0.00 & 0.00 & 10.34 & 89.66 & -0.01 & 9.72 & 9.95 & 0.0276 & 0.02202 & 2.83 \\
1-285031 &  0.00 & 9.10 & 0.00 & 0.00 & 18.70 & 0.00 & 72.37 & 0.12 & 0.00 & 0.00 & 3.13 & 0.00 & 96.75 & 0.88 & 9.57 & 10.07 & 0.01309 & 0.01318 & 2.50 \\
1-285052 &  0.00 & 1.84 & 0.00 & 0.00 & 6.78 & 0.00 & 87.30 & 0.01 & 0.00 & 0.00 & 1.06 & 0.00 & 98.92 & 0.12 & 9.95 & 10.10 & 0.0202 & 0.01998 & 3.31 \smallskip \\
1-210785 &  0.00 & 2.36 & 0.00 & 0.00 & 5.12 & 7.49 & 81.03 & 0.01 & 0.00 & 0.00 & 0.83 & 1.62 & 97.54 & 0.00 & 9.91 & 10.09 & 0.01956 & 0.02166 & 1.97 \\
1-286804 &  0.00 & 0.00 & 8.24 & 16.66 & 24.32 & 42.92 & 5.58 & 0.00 & 0.79 & 5.34 & 20.46 & 48.78 & 24.63 & 0.33 & 8.95 & 9.34 & 0.02839 & 0.02756 & 3.96 \\
1-289865 &  2.24 & 0.00 & 0.00 & 0.00 & 0.00 & 0.00 & 94.47 & 0.00 & 0.00 & 0.00 & 0.00 & 0.00 & 100.00 & -0.21 & 10.08 & 10.10 & 0.02269 & 0.02628 & 1.70 \\
1-295095 &  2.49 & 0.00 & 0.00 & 0.00 & 0.00 & 37.87 & 56.60 & 0.00 & 0.00 & 0.00 & 0.00 & 14.38 & 85.62 & 0.07 & 9.69 & 9.95 & 0.0273 & 0.0198 & 3.53 \\
1-320681 &  0.00 & 0.00 & 0.00 & 0.00 & 0.00 & 10.53 & 85.98 & 0.00 & 0.00 & 0.00 & 0.00 & 1.25 & 98.75 & -0.07 & 10.01 & 10.10 & 0.02525 & 0.02942 & 1.70 \smallskip \\
1-338828 &  0.00 & 16.73 & 7.73 & 0.00 & 8.65 & 58.85 & 7.16 & 0.43 & 0.44 & 0.00 & 5.37 & 51.09 & 42.67 & 1.00 & 8.78 & 9.58 & 0.02418 & 0.03773 & 4.30 \\
1-339028 &  2.72 & 0.00 & 0.00 & 0.00 & 0.00 & 0.00 & 94.05 & 0.00 & 0.00 & 0.00 & 0.00 & 0.00 & 100.00 & -0.08 & 10.11 & 10.11 & 0.02326 & 0.02611 & 2.33 \\
1-339125 &  0.00 & 1.99 & 0.00 & 0.00 & 0.00 & 0.57 & 94.78 & 0.01 & 0.00 & 0.00 & 0.00 & 0.07 & 99.92 & 0.41 & 10.02 & 10.11 & 0.01708 & 0.02065 & 3.18 \\
1-351538 &  7.68 & 2.73 & 0.00 & 0.00 & 6.55 & 26.05 & 53.68 & 0.05 & 0.00 & 0.00 & 1.18 & 7.80 & 90.97 & 0.32 & 9.65 & 10.03 & 0.02409 & 0.0241 & 3.94 \\
1-36878 &  0.00 & 8.72 & 0.00 & 2.05 & 13.83 & 45.19 & 27.80 & 0.17 & 0.00 & 0.26 & 4.55 & 19.97 & 75.05 & 0.84 & 9.25 & 9.90 & 0.01882 & 0.02383 & 2.21 \smallskip \\
1-37062 &  1.96 & 5.04 & 18.39 & 0.00 & 32.06 & 16.55 & 23.61 & 0.08 & 0.70 & 0.00 & 12.73 & 8.76 & 77.73 & 0.77 & 8.88 & 9.88 & 0.02567 & 0.02812 & 2.35 \\
1-37079 &  15.94 & 4.62 & 0.00 & 0.00 & 0.00 & 59.29 & 17.37 & 0.13 & 0.00 & 0.00 & 0.00 & 57.50 & 42.37 & 0.06 & 9.25 & 9.46 & 0.02414 & 0.02946 & 7.75 \\
1-377125 &  12.64 & 0.00 & 0.00 & 0.00 & 0.00 & 31.51 & 53.78 & 0.00 & 0.00 & 0.00 & 0.00 & 10.73 & 89.27 & 0.16 & 9.77 & 10.01 & 0.03576 & 0.03091 & 4.22 \\
1-377321 &  21.35 & 27.63 & 0.00 & 15.90 & 0.00 & 25.58 & 7.92 & 1.34 & 0.00 & 5.66 & 0.00 & 25.86 & 67.14 & 0.20 & 8.28 & 9.71 & 0.01203 & 0.03579 & 2.33 \\
1-378401 &  0.00 & 0.00 & 0.00 & 1.58 & 0.00 & 0.00 & 95.58 & 0.00 & 0.00 & 0.08 & 0.00 & 0.00 & 99.92 & -0.05 & 10.09 & 10.11 & 0.02769 & 0.03087 & 2.46 \smallskip \\
1-378795 &  1.69 & 0.00 & 0.00 & 0.00 & 0.00 & 12.23 & 81.65 & 0.00 & 0.00 & 0.00 & 0.00 & 3.13 & 96.87 & 0.24 & 9.93 & 10.06 & 0.02149 & 0.02 & 3.81 \\
1-379087 &  15.45 & 5.06 & 3.25 & 0.00 & 4.42 & 21.55 & 47.54 & 0.06 & 0.08 & 0.00 & 1.19 & 7.48 & 91.19 & 0.37 & 9.51 & 10.02 & 0.01631 & 0.02329 & 2.87 \\
1-379660 &  0.00 & 1.11 & 0.00 & 0.00 & 36.10 & 12.52 & 47.77 & 0.02 & 0.00 & 0.00 & 9.50 & 4.45 & 86.03 & 0.45 & 9.50 & 9.94 & 0.02892 & 0.02331 & 1.66 \\
1-386452 &  0.00 & 5.48 & 0.00 & 2.46 & 7.83 & 40.57 & 41.26 & 0.11 & 0.00 & 0.23 & 1.97 & 13.57 & 84.12 & 0.30 & 9.45 & 9.98 & 0.0168 & 0.02326 & 1.60 \\
\hline
\end{tabular}
\end{table}
\end{landscape}

\begin{landscape}
\begin{table}
\contcaption{}
\begin{tabular}{ccccccccccccccccccccc}
\hline
mangaID & FC  & xyy & xyo & xIy & xII & xIo & xo  &  myy & myo  & mIy  & mII  & mIo  & mo   & $A_V$ & $\langle t_L\rangle$ & $\langle t_M\rangle$ & $\langle Z_L\rangle$ & $\langle Z_M\rangle$ & Adev  \\
 (1)    & (2) & (3) & (4) & (5) & (6) & (7) & (8) & (9)  & (10) & (11) & (12) & (13) & (14) & (15)  & (16)    & (17)    & (18)    & (19)    & (20)  \\
\hline
1-386695 &  19.61 & 24.78 & 0.00 & 9.17 & 0.00 & 30.32 & 15.10 & 0.78 & 0.00 & 2.01 & 0.00 & 17.65 & 79.55 & 0.14 & 8.55 & 9.89 & 0.01586 & 0.03372 & 3.29 \\
1-392976 &  1.90 & 0.00 & 0.00 & 0.00 & 0.00 & 8.78 & 87.13 & 0.00 & 0.00 & 0.00 & 0.00 & 2.95 & 97.05 & 0.08 & 9.93 & 10.04 & 0.01636 & 0.01699 & 3.38 \\
1-322671 &  0.00 & 0.00 & 0.00 & 0.00 & 0.00 & 32.03 & 63.35 & 0.00 & 0.00 & 0.00 & 0.00 & 9.53 & 90.47 & 0.00 & 9.76 & 9.99 & 0.02973 & 0.0227 & 2.53 \\
1-43009 &  9.00 & 0.00 & 0.00 & 0.00 & 0.00 & 37.60 & 52.89 & 0.00 & 0.00 & 0.00 & 0.00 & 14.31 & 85.69 & 0.31 & 9.68 & 9.90 & 0.02034 & 0.02292 & 4.45 \\
1-43721 &  4.01 & 0.00 & 0.00 & 0.00 & 0.00 & 0.00 & 94.49 & 0.00 & 0.00 & 0.00 & 0.00 & 0.00 & 100.00 & 0.14 & 10.11 & 10.11 & 0.02641 & 0.03012 & 3.26 \smallskip \\
1-44789 &  0.00 & 0.00 & 0.00 & 0.00 & 0.00 & 0.00 & 96.11 & 0.00 & 0.00 & 0.00 & 0.00 & 0.00 & 100.00 & 0.17 & 10.08 & 10.10 & 0.01943 & 0.02304 & 3.57 \\
1-47499 &  0.00 & 1.90 & 0.00 & 0.00 & 11.79 & 0.00 & 83.53 & 0.03 & 0.00 & 0.00 & 2.29 & 0.00 & 97.68 & 0.39 & 9.78 & 10.04 & 0.01892 & 0.01871 & 2.97 \\
1-48208 &  0.00 & 1.62 & 0.00 & 0.00 & 0.00 & 0.00 & 94.83 & 0.02 & 0.00 & 0.00 & 0.00 & 0.00 & 99.98 & -0.11 & 10.04 & 10.11 & 0.02153 & 0.0236 & 2.06 \\
1-487130 &  0.00 & 10.02 & 0.00 & 0.00 & 8.67 & 51.80 & 27.75 & 0.17 & 0.00 & 0.00 & 2.48 & 25.46 & 71.89 & 0.42 & 9.19 & 9.86 & 0.01724 & 0.02632 & 3.97 \\
1-489649 &  0.00 & 0.84 & 0.00 & 0.00 & 0.00 & 0.00 & 95.53 & 0.01 & 0.00 & 0.00 & 0.00 & 0.00 & 99.99 & 0.21 & 9.99 & 10.08 & 0.01646 & 0.01877 & 2.75 \smallskip \\
1-491233 &  0.00 & 4.43 & 3.77 & 3.32 & 1.99 & 39.65 & 43.89 & 0.07 & 0.09 & 0.32 & 0.51 & 13.43 & 85.59 & 0.60 & 9.44 & 9.99 & 0.01447 & 0.02146 & 2.67 \\
1-519738 &  0.00 & 0.00 & 0.00 & 0.00 & 0.00 & 15.67 & 79.93 & 0.00 & 0.00 & 0.00 & 0.00 & 3.51 & 96.49 & -0.09 & 9.97 & 10.08 & 0.02318 & 0.02071 & 1.74 \\
1-52259 &  3.49 & 5.53 & 0.00 & 0.00 & 29.55 & 32.81 & 26.06 & 0.17 & 0.00 & 0.00 & 10.76 & 16.19 & 72.88 & 0.43 & 9.22 & 9.82 & 0.02862 & 0.03064 & 3.23 \\
1-55572 &  0.00 & 1.96 & 0.00 & 1.96 & 0.00 & 0.00 & 93.04 & 0.02 & 0.00 & 0.10 & 0.00 & 0.00 & 99.88 & -0.22 & 10.00 & 10.11 & 0.02372 & 0.0266 & 1.87  \\
1-604048 &  0.00 & 2.13 & 0.00 & 0.00 & 18.87 & 1.09 & 77.52 & 0.03 & 0.00 & 0.00 & 3.19 & 0.30 & 96.48 & 0.15 & 9.78 & 10.06 & 0.01837 & 0.01727 & 1.70 \smallskip\\
1-613211 &  0.00 & 1.37 & 0.00 & 0.33 & 0.00 & 0.00 & 94.87 & 0.01 & 0.00 & 0.02 & 0.00 & 0.00 & 99.97 & -0.04 & 10.05 & 10.11 & 0.02866 & 0.03107 & 2.14 \\
1-626830 &  0.00 & 1.90 & 0.00 & 3.56 & 26.07 & 10.95 & 55.32 & 0.02 & 0.00 & 0.34 & 6.63 & 3.71 & 89.28 & 0.55 & 9.51 & 9.96 & 0.02513 & 0.02435 & 2.59 \\
1-322074 &  0.00 & 0.00 & 0.00 & 0.00 & 0.00 & 39.98 & 56.43 & 0.00 & 0.00 & 0.00 & 0.00 & 13.89 & 86.11 & 0.08 & 9.72 & 9.98 & 0.02835 & 0.01969 & 2.49  \\
1-633990 &  7.36 & 22.54 & 0.00 & 0.00 & 0.00 & 37.91 & 30.08 & 0.29 & 0.00 & 0.00 & 0.00 & 17.12 & 82.58 & 0.56 & 9.04 & 9.98 & 0.01428 & 0.02867 & 2.98 \\
1-633994 &  0.00 & 1.49 & 0.00 & 0.00 & 10.82 & 9.43 & 75.39 & 0.01 & 0.00 & 0.00 & 2.01 & 2.57 & 95.41 & 1.19 & 9.80 & 10.05 & 0.0248 & 0.02387 & 3.11 \smallskip\\
1-48053 &  0.00 & 4.42 & 0.00 & 0.00 & 0.00 & 0.00 & 92.42 & 0.03 & 0.00 & 0.00 & 0.00 & 0.00 & 99.97 & 0.00 & 9.96 & 10.11 & 0.02649 & 0.02932 & 1.76 \\
1-635503 &  0.00 & 22.47 & 10.86 & 0.00 & 0.00 & 35.45 & 32.37 & 0.39 & 0.41 & 0.00 & 0.00 & 19.28 & 79.93 & 1.15 & 8.78 & 9.89 & 0.01045 & 0.02839 & 3.25 \\
1-71481 &  3.01 & 0.00 & 0.00 & 0.00 & 0.00 & 2.51 & 91.76 & 0.00 & 0.00 & 0.00 & 0.00 & 0.28 & 99.72 & -0.07 & 10.09 & 10.11 & 0.0274 & 0.02989 & 2.22 \\
1-71525 &  0.00 & 2.20 & 0.00 & 0.00 & 0.00 & 0.00 & 94.16 & 0.02 & 0.00 & 0.00 & 0.00 & 0.00 & 99.98 & 0.33 & 9.87 & 10.05 & 0.01612 & 0.02106 & 2.99 \\
1-72914 &  5.38 & 0.00 & 0.00 & 0.00 & 0.00 & 23.39 & 68.11 & 0.00 & 0.00 & 0.00 & 0.00 & 5.85 & 94.15 & 0.19 & 9.90 & 10.07 & 0.02155 & 0.02217 & 3.35 \smallskip  \\
1-72928 &  0.00 & 0.00 & 0.00 & 0.00 & 0.00 & 0.00 & 96.34 & 0.00 & 0.00 & 0.00 & 0.00 & 0.00 & 100.00 & -0.14 & 10.06 & 10.10 & 0.02221 & 0.02526 & 2.31 \\
1-73005 &  0.00 & 0.00 & 0.00 & 0.00 & 17.26 & 39.48 & 41.22 & 0.00 & 0.00 & 0.00 & 4.85 & 14.80 & 80.35 & 0.32 & 9.60 & 9.93 & 0.01782 & 0.01843 & 2.85 \\
1-90849 &  4.99 & 0.00 & 0.00 & 0.00 & 18.92 & 1.51 & 70.51 & 0.00 & 0.00 & 0.00 & 3.84 & 0.45 & 95.70 & 0.24 & 9.81 & 10.05 & 0.02369 & 0.01937 & 3.30 \\
1-92626 &  1.70 & 0.65 & 0.00 & 0.00 & 0.00 & 0.00 & 92.91 & 0.01 & 0.00 & 0.00 & 0.00 & 0.00 & 99.99 & 0.03 & 10.04 & 10.10 & 0.02273 & 0.02675 & 2.69 \\
1-93876 &  4.73 & 0.00 & 0.00 & 0.00 & 0.00 & 20.53 & 71.66 & 0.00 & 0.00 & 0.00 & 0.00 & 5.07 & 94.93 & -0.08 & 9.89 & 10.05 & 0.02999 & 0.02718 & 3.64 \smallskip  \\
1-94422 &  3.24 & 0.00 & 0.00 & 0.00 & 0.00 & 25.12 & 67.74 & 0.00 & 0.00 & 0.00 & 0.00 & 6.84 & 93.16 & 0.17 & 9.83 & 10.04 & 0.02905 & 0.0258 & 2.86 \\
1-94554 &  4.48 & 0.00 & 0.00 & 0.00 & 0.00 & 5.49 & 86.38 & 0.00 & 0.00 & 0.00 & 0.00 & 1.08 & 98.92 & 0.01 & 10.07 & 10.11 & 0.02174 & 0.02448 & 2.25 \\
\hline
\end{tabular}
\end{table}
\end{landscape}


\bsp	
\label{lastpage}
\end{document}